\pdfoutput=1
\documentclass[10pt, logo, copyright, nonumbering]{nvidiatechreport}
\usepackage{cleveref}
\usepackage[numbers,sort&compress]{natbib}
\usepackage[version=4]{mhchem}
\usepackage{placeins}
\usepackage{subcaption}
\usepackage{tcolorbox}
\newsavebox{\reactionkeybox}
\usepackage{physics}

\usepackage{xcolor}
\hypersetup{
  colorlinks=true,
  urlcolor=blue,
  citecolor=blue,
  linkcolor=blue
}

\providecommand{\doi}[1]{}
\renewcommand{\doi}[1]{DOI: %
  \href{https://doi.org/#1}{\nolinkurl{#1}}%
}
\newcommand{\isbn}[1]{ISBN: %
  \href{https://search.worldcat.org/isbn/#1}{\nolinkurl{#1}}%
}

\title{\centering{Reducing $\text{T}$ gates in product formulas\\ with agent loops}}
\author{
{\centering
Luis {Mantilla Calder\'on}, J\'er\^ome F.\ Gonthier, Alex Zook, Kohei Nakaji, Taylor L. Patti, 
Al\'an Aspuru-Guzik\par
\vspace{0.75em}
{\normalfont\mdseries\footnotesize
\setlength{\parskip}{0pt}
\setlength{\parindent}{0pt}
NVIDIA Corporation, Santa Clara, CA, USA\par

}
}
}

\begin{abstract}
\textbf{Abstract:} Quantum chemistry calculations on a fault-tolerant quantum computer usually require many $\text{T}$ gates. This cost depends on the simulation method and how precisely its rotations are implemented. In this paper, we used Nemotron 3.5 Lightning, a large language model (LLM), in a coding-agent loop to optimize the structure and rotation-synthesis precision of product formulas (PFs) for quantum phase estimation (QPE). Each of five independent searches optimized a shared protocol across 115 HamLib molecular Hamiltonians spanning 4--16 qubits over 100 code mutations. The searches started from a calibrated, nested second-order PF with uniform synthesis precision. We minimized $T_{95}$, the estimated total $\text{T}$-gate cost including QPE queries and repeated attempts for 95\% success probability, while requiring chemical accuracy for every training Hamiltonian. We name the best protocol found error-limited optimized Trotter evolution (ELOTE). It achieves chemical accuracy for its training data and achieved median and geometric-mean $T_{95}$ reductions of $85.35\%$ and $84.82\%$, respectively. ELOTE combines contributions to the same Pauli string, uses a first-order PF, and varies synthesis precision with coefficient magnitude, Pauli support, and diagonal character. We also tested ELOTE on 76 larger Hamiltonians with 18--22 qubits, which met chemical accuracy in every case. For ten reaction energies, we further optimized a shared $\delta$-ELOTE protocol. It reduced geometric-mean $T_{95}$ by $58.90\%$ relative to ELOTE while preserving chemical accuracy for every signed reaction difference. These classically verified results show cost savings from shared protocols across molecules and reactions, and motivate similar searches in which energy measurements from quantum computers guide explorations in a coding-agent loop.
\end{abstract}

\begin{document}

\maketitle
\suppressfloats[t]

\section{Introduction}

Hamiltonian simulation underlies many proposed fault-tolerant quantum algorithms for chemistry~\cite{aspuruguzik2005simulated,jones2012faster,reiher2017reaction,vonburg2021catalysis,goings2022cytochrome,schleich2025quantum}. Their resource requirements often depend strongly on non-Clifford gates, making the $\text{T}$ or Toffoli count a central optimization target~\cite{simon2025resource,cain2026shor}. Product formulas (PFs) implement approximate evolution $e^{\mathrm{i}tH}$ through ordered products of Pauli rotations~\cite{trotter1959,suzuki1990,suzuki1992,bagherimehrab2024cpf}. Alternative constructions, including linear combinations of unitaries~\cite{berry2015taylor}, quantum singular value transformation~\cite{gilyen2019qsvt}, and qubitization~\cite{low2019qubitization,berry2019qubitizationchem}, provide other routes to implementing the exponential of the Hamiltonian. In chemistry, these simulation methods can be used to estimate the energy of an Eigenstate of the underlying Hamiltonian~\cite{aspuruguzik2005simulated}. In this work, we aim to reduce the resources needed for energy estimation protocols that make use of a Hamiltonian simulation subroutine. In particular, we focus on PFs since this method does not require ancilla quibts, and even though asymptotic performance is worse than qubitization~\cite{low2019qubitization}---in practice the simulation error tends to be lower than what some bounds suggest~\cite{reiher2017reaction, childs2021trottererror}. We then optimize PFs when used within quantum phase estimation (QPE) and reduce the $\text{T}$ gate cost while preserving chemical accuracy for the target energy.

The error of energy estimates obtained with QPE can be broken down into three categories: phase readout precision $(\epsilon_{\mathrm{QPE}})$, Hamiltonian simulation approximation $(\epsilon_{\mathrm{sim}})$, and finite-precision synthesis of the rotations or arithmetic used in the circuit $(\epsilon_{\rm synth})$. The success probability of QPE also depends on the input state, however this is not an inherit QPE error and thus it is not included in the error decomposition. Therefore, for the QPE routine to be successful in a quantum chemistry setting, the difference between the QPE energy estimate $\widehat E_{\mathcal P}$ and the exact value $E_0$ should be within chemical accuracy ($\varepsilon_{\rm chem}=1\ \mathrm{kcal/mol} \simeq 1.6 \mathrm{mHa}$),
\[
|\widehat E_{\mathcal P}-E_0|
\lesssim
\epsilon_{\mathrm{QPE}}
+ \epsilon_{\mathrm{sim}}
+ \epsilon_{\mathrm{synth}}
\le \varepsilon_{\rm chem}.
\]

The allocation of this budget creates a resource trade-off:
reducing the phase-readout error leaves more room for simulation
and synthesis errors, but instead requires a larger amount of controlled evolutions. For example, recent catalyst resource estimates assign fixed fractions for these three terms~\cite{bellonzi2024catalyst}. Any allocation of these three terms will have an impact in the cost of T gates needed to run an algorithm: a smaller synthesis precision $\epsilon_{\mathrm{synth}}$ will require using more T gates when expanding single qubit rotations, or a smaller simulation error $\epsilon_{\mathrm{sim}}$ could require a higher order PF leading to more Pauli rotations and thus the need for more T gates.
Even after these budgets are fixed, there is still freedom in how the product formula is constructed and how synthesis tolerances are distributed among its individual rotations. Different rotations need not affect the target energy equally, and their effects depend on the splitting, ordering, and synthesis of the complete evolution. We hypothesize that jointly adapting these choices can reduce the total $\text{T}$-gate cost of energy estimation relative to a uniform-precision baseline while preserving chemical accuracy across an ensemble of molecular Hamiltonians.

In this work, we use a large language model (LLM) in a coding-agent loop to progressively optimize the structure and rotation synthesis
tolerances for a PF, while being transferable across different molecular Hamiltonians. We minimize the total $\text{T}$-gate cost of energy
estimation used for QPE, taking into consideration the amount of QPE queries and repeated attempts needed to reach chemical accuracy. We do this optimization with a subset of chemical Hamiltonians from HamLib~\cite{sawaya2024hamlib}, using 115 small Hamiltonians for guiding the discovery loop and 76 larger-sized Hamiltonians for testing the resulting protocol. We call the best protocol found error-limited optimized Trotter evolution (ELOTE). This protocol combines a modified splitting with coefficient aggregation and nonuniform synthesis tolerances. We then adapt ELOTE to ten reaction energies, optimizing a shared protocol we call $\delta$-ELOTE, which can further reduce $\text{T}$ gate count when the constraint is an energy difference rather than an absolute energy estimation. The current paper is divided as follows: \Cref{sec:relwork} reviews chemistry-tailored quantum simulation and frameworks for agentic loops. \Cref{sec:methods} formulates the problem we target, and \Cref{sec:experiments} presents the molecule-energy and reaction-energy optimization results. Appendix~\ref{app:evaluator} describes the way we benchmarked the protocols and Appendix~\ref{app:understanding_elote} explains the best performing protocols that were found in our work.
\section{Related Work}
\label{sec:relwork}

\subsection{Chemistry-tailored quantum simulation}

Quantum algorithms for chemistry can reduce their resource requirements by exploiting the structure of electronic Hamiltonians. Sparsity and low-rank factorization enable more efficient qubitized simulation~\cite{berry2019qubitizationchem}. Double factorization~\cite{vonburg2021catalysis}, tensor hypercontraction~\cite{lee2021thc}, and other factorization methods~\cite{king2026quantum} further exploit structure in the two-electron integrals to reduce the cost of Hamiltonian block encodings. For PFs, chemical structure also influences approximation error: nuclear charge and orbital filling can be more informative than orbital count alone in explaining ground-state energy errors~\cite{babbush2015chemical}. These results motivate adapting simulation protocols to both the Hamiltonian and the energy being estimated.

Another place to look for structure that can be exploited for chemistry simulations is the precision of rotation synthesis. In fault-tolerant circuits, arbitrary-angle rotations are
approximated by sequences of Clifford+$\text{T}$ gates (or similar universal gate sets) whose synthesis cost
grows asymptotically logarithmically with inverse
tolerance~\cite{kliuchnikov2013,rossselinger2016}.
Resource estimates often assign a uniform tolerance to these
rotations. For example, in Ref.~\cite{babbush2018linear}, the
synthesis budget is divided evenly among all $z$
rotations. Other methods, such as Ref.~\cite{bothe2026smallangle}. propose angle-aware synthesis, where small rotations are exploited and synthesized probabilistically~\cite{bothe2026smallangle}.
The resulting resource savings coming just from better rotation synthesis methods are quite significant, and substantially benefit quantum simulation.

Mixed-precision algorithms allocate numerical accuracy according to the requirements of different computational steps~\cite{highammary2022}, with applications in neural-network training~\cite{micikevicius2018mixed} and classical quantum chemistry. For example, mixed-precision density matrix renormalization group (DMRG) calculations use the Ozaki scheme to emulate 64-bit floating-point (FP64) arithmetic with 8-bit integer (INT8) slices on Blackwell graphics processing units (GPUs), maintaining milli-Hartree agreement with native FP64 results~\cite{brower2025blackwell}. Blackwell Tensor Cores also support NVFP4, a block-scaled four-bit floating-point format~\cite{nvidia2025nvfp4}. For fault-tolerant quantum simulation, the corresponding goal is to assign rotation-synthesis tolerances according to their effect on the target observable. Joint algorithm and hardware design could additionally consider magic-state factories with different output error rates.

Here, we investigate how to optimize resources in PFs by a joint optimization of the splitting, Hamiltonian term grouping, ordering, repetitions, and
synthesis tolerances methods--optimized within a subset of molecular Hamiltonias from HamLib~\cite{sawaya2024hamlib}. Each choice affects both the compiled evolution and the precision budget available for QPE readout.Thus, the protocol aims to optimize the total $\text{T}$-gate cost spent for measuring the ground-state energy from a ground state input using QPE.

\subsection{Discovery with agentic loops}
\label{sec:searchloops}
Recently, coding agents have been used for scientific discovery, tackling very complex problems such as the Euler and Navier--Stokes singularities~\cite{alpoge2026forcedeuler,openai2026navierstokes}, the Exponential Time Hypothesis, and existence questions for quantum chromodynamics (QCD) and Yang--Mills theories~\cite{openai2026scientificunderstanding}. Works based on agentic loops have also improved mathematical constructions and quantum circuits by repeatedly proposing candidates and evaluating them against explicit objectives~\cite{georgiev2026mathematicaldiscovery,gu2026quantumcircuits}. This idea of updating code, evaluating it, and using some feedback to improve candidate solutions, has been used in the past in multiple different ways. Some initial works include: FunSearch~\cite{romeraparedes2024funsearch} evolving functions for mathematical and algorithmic problems and AlphaEvolve~\cite{novikov2025alphaevolve, openevolve2025} extending this approach to broader code optimization. Related methods search over prompts and heuristics~\cite{chen2023evoprompting,liu2024eoh} or adapt candidate generation and selection~\cite{lange2025shinkaevolve,ye2024reevo,yu2025satlution}.

Other approaches use reinforcement learning (RL) for tuning the base model toward a specific goal, learning how to generate better candidates from rewards. AlphaTensor~\cite{fawzi2022alphatensor} and AlphaDev~\cite{mankowitz2023alphadev} trained policies to discover matrix-multiplication and sorting algorithms. TTT-Discover~\cite{yuksekgonul2026tttdiscover} applied RL during the solution of a test problem, updating the LLM using experience specific to that problem. Search feedback therefore changes the proposal policy itself, allowing subsequent attempts to draw on knowledge encoded in updated model parameters as well as the recorded history of the search.

A different research direction aims to progressively modify the agent's harness. The Darwin G\"odel Machine~\cite{zhang2025dgm} evolves agents that edit their own code, and Hyperagents~\cite{zhang2026hyperagents} also make the self-modification mechanism editable. EvoX~\cite{liu2026evox} adapts the search strategies used to generate candidates. At the level of prompts, GEPA~\cite{agrawal2025gepa} uses reflection on execution traces to guide evolution. SoL-Pi~\cite{liu2026solpi} searches for improvements in tool execution and context handling. These approaches optimize the machinery that conducts the search, including how evidence is presented and how proposed changes are generated and tested.

In chemistry and quantum simulation, agent workflows have been used to organize molecular knowledge and solve computational tasks~\cite{gustin2025elagente,zou2025elagente,pham2026chemgraph,fu2026qagent}. When used in an agent loop, they can also optimize the performance of a quantum algorithm~\cite{mantilla2026autoresearch,finger2026automated}.
Agent-based resource optimization has also been applied to quantum circuits for quantum simulation, but has been limited to individual molecular species~\cite{gu2026quantumcircuits,zhang2025moleculargeometry}. Our work uses agent loops to optimize a PF protocol shared across multiple molecular Hamiltonians. The search jointly changes the evolution sequence and synthesis tolerances, retaining protocols that reduce QPE-inclusive $\text{T}$-gate cost while preserving chemical accuracy for every Hamiltonian in the ensemble.

\section{Methods}
\label{sec:methods}

\begin{figure}[t]
\centering
\includegraphics[width=\linewidth]{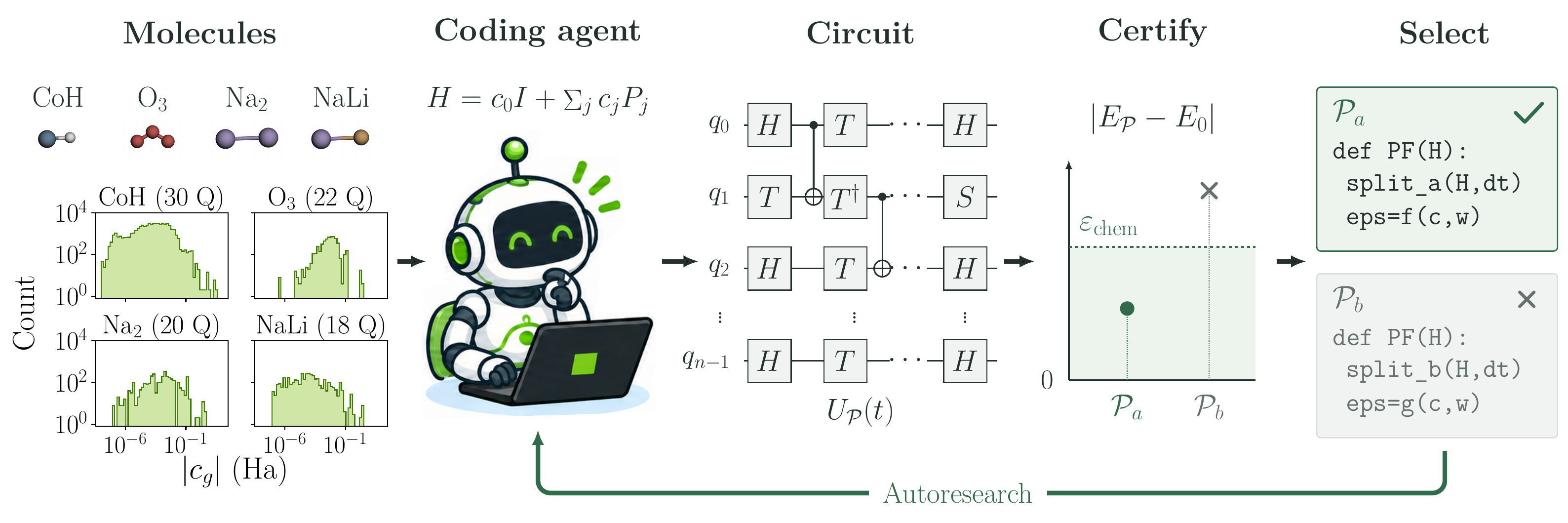}
\caption{A schematic of the ideal agentic loop used to optimize Hamiltonian simulation. Given a training set of molecules, the coding agent proposes ways to optimize a protocol that can be verified with quantum measurement. For this loop to be efficient on a quantum computer, one must design a fast and accurate energy-certification procedure as the proxy that guides the agentic loop.}
\label{fig:autoresearch_loop}
\end{figure}

Here, we discuss our approach for optimizing and benchmarking a coding agent that modifies the structure and synthesis tolerances of a PF protocol, which is shared across a training ensemble of Hamiltonians (Figure~\ref{fig:autoresearch_loop}). The search starts from a baseline PF, we chose it to be that of Ref.~\cite{reiher2017reaction}. Then, energy estimates and gate counts from each compiled proposal guide subsequent mutations. We retain lower-cost protocols that satisfy an energy constraint imposed for all Hamiltonians in a training dataset. We used Nemotron 3.5 Lightning~\cite{nvidia2026nemotron35} to propose code changes, OpenCode~\cite{opencode_ai} as the coding-agent harness, and OpenShell~\cite{nvidia_openshell_2026} to sandbox execution during the mutation loop.

On quantum hardware, energy feedback could come from QPE outcomes~\cite{nielsen2010quantum} or energy-estimation and certification methods~\cite{lin2022heisenberg,wang2025energycertification} (Appendix~\ref{app:energy_witness}), without full tomography. Such tests require reference information and controlled statistical uncertainty. State fidelity alone does not guarantee accuracy in energy estimates from QPE. Here, we evaluate the compiled evolution classically against reference eigenpairs while counting the QPE-inclusive circuit cost. The Hamiltonians, evolution times, input states, and cost definition remain fixed across proposals.

\subsection{Objective: Minimize \texorpdfstring{\(\text{T}\)}{T}-gates}

A PF can be abstracted into a protocol \(\mathcal P\) which defines one step of the evolution as a sequence of \(L\) Pauli rotations. It specifies how many times to repeat this step, \(r\), and how accurately to synthesize each rotation, \(\epsilon_j\). Omitting the identity phase, the evolution before gate synthesis is
\begin{equation}
U_{\mathcal P}(t)
=
\left( \prod_{j=1}^{L}
R_j\!\left(\omega_j\right) \right)^{r},
\label{eq:mixed_precision_pf_unitary}
\end{equation}
where \(R_j(\omega_j)=e^{\mathrm{i}\omega_jP_j}\). The angle \(\omega_j\) depends on the Hamiltonian coefficient, time step, and splitting. The same Pauli string \(P_j\) can appear more than once. Grouping and ordering the terms, or combining contributions to the same string, changes the sequence of rotations. We evaluate the resulting circuit by its \(\text{T}\)-gate cost and the error in the energy it encodes.

\paragraph{\texorpdfstring{\(\text{T}\)}{T}-gate costs.}
We implement each Pauli rotation by first applying Clifford gates \(C_j\) chosen for the string \(P_j\). We then rotate one qubit about the \(z\) axis and undo the Clifford gates~\cite{nielsen2010quantum}. With \(R_z(\theta)=e^{-i\theta Z/2}\),
\begin{equation}
e^{i\omega_jP_j}
=
C_j^\dagger R_z(-2\omega_j) C_j.
\label{eq:pauli_to_rz_main}
\end{equation}
The Clifford gates contribute no \(\text{T}\) gates. We approximate the single-qubit rotation with a sequence of Clifford and \(\text{T}\) gates using Ross--Selinger synthesis in CUDA-Q~\cite{rossselinger2016,nvidia_cudaq_rotation_synthesis}. Counting the \(\text{T}\) gates in every rotation and all \(r\) repetitions gives
\begin{equation}
N_T
=
r\sum_{j=1}^{L}
\tau\!\left(-2\omega_j,\epsilon_j\right).
\label{eq:main_pf_t_cost}
\end{equation}
Here \(\tau(\theta,\epsilon)\) counts the \(\text{T}\) gates in the compiled approximation to \(R_z(\theta)\) at tolerance \(\epsilon\). Thus, \(N_T\) is the cost of one full evolution. Assuming the input state is the exact ground state of \(H\), our objective, \(T_{95}\), estimates the cost of enough QPE attempts to obtain at least one energy estimate within chemical accuracy of \(E_0\) with probability at least \(95\%\), charging \(N_T\) per evolution call. A larger compiled energy error leaves less tolerance for QPE readout, which increases the number of circuit calls in the cost model. A less probable energy outcome leads to more attempts. See Appendix~\ref{app:tgate_cost_derivations} for details on the cost calculation.

\paragraph{Chemical-accuracy constraint.}
QPE infers energy from the phase acquired during evolution. Let \(E\) be the reference energy, \(E_{\mathcal P}\) the energy encoded by the selected eigenphase of the compiled circuit, and \(\widehat E_{\mathcal P}\) its finite-precision QPE estimate. On a successful readout with tolerance \(\epsilon_{\rm QPE}\),
\[
|\widehat E_{\mathcal P}-E|
\le |E_{\mathcal P}-E|+|\widehat E_{\mathcal P}-E_{\mathcal P}|
\le |E_{\mathcal P}-E|+\epsilon_{\rm QPE}.
\]
For every molecule and reaction species, the input is its exact ground state with the prescribed electron number. This fixes the state preparation so that the compiled energy bias \(E_{\mathcal P}-E\) comes from the PF and gate synthesis. We minimize cost among protocols whose compiled energy error leaves a positive budget for QPE readout:
\begin{equation}
\mathcal F
:=
\left\{
\mathcal P:
\left|E_{\mathcal P}-E\right|
<
\varepsilon_{\rm chem}
\right\},
\qquad
\mathcal P^{\star}
\in
\operatorname*{arg\,min}_{\mathcal P\in\mathcal F}
T_{95}(\mathcal P),
\label{eq:observable_feasible_set}
\end{equation}

Every valid protocol must satisfy this accuracy requirement for every Hamiltonian in the training ensemble \(\mathcal D\). To compare costs across molecules, we divide each \(T_{95}\) by its tight-baseline value and minimize the geometric mean of these ratios:
\begin{equation}
\mathcal L(\mathcal P)
=\exp\!\left[
\frac{1}{|\mathcal D|}\sum_{H\in\mathcal D}
\log\frac{T_{95}(H,\mathcal P)}{T_{95}^{(0)}(H)}
\right],
\qquad
|E_{H,\mathcal P}-E_H|<\varepsilon_{\rm chem}
\quad\text{for every }H\in\mathcal D.
\label{eq:shared_protocol_objective}
\end{equation}
$T_{95}^{(0)}(H)$ denotes the $\text{T}$-cost for tight baseline. A proposal is retained when it lowers \(\mathcal L\), yields numerically converged energy and QPE success-probability estimates, and satisfies the accuracy constraint for every Hamiltonian. This objective allows a cost increase for some Hamiltonians if the geometric mean decreases.

For a reaction, the target is the difference between the total product and reactant energies, including stoichiometric weights. Write \(\Delta E=E_{\rm p}-E_{\rm r}\) and \(\Delta E_{\mathcal P}=E_{{\rm p},\mathcal P}-E_{{\rm r},\mathcal P}\) for the reference and compiled differences. Their error obeys
\[
\left|\Delta E_{\mathcal P}-\Delta E\right|
\le
\left|E_{{\rm p},\mathcal P}-E_{\rm p}\right|
+
\left|E_{{\rm r},\mathcal P}-E_{\rm r}\right|.
\]
We impose chemical accuracy on the reaction-energy error on the left. Errors in the individual species can therefore cancel as discussed in Ref.~\cite{baysmidt2026nanographenes}. Appendix~\ref{app:energy_evaluation} explains how we select the energy outcome and assign QPE precision to each species.

\subsection{Training data and baseline}

To test and optimize the performance of a PF, we restrict it to an ensemble \(\mathcal D\) of Hamiltonians that we can test in some hardware, either classical or quantum. For molecular simulation, this ensemble can span different chemical species, geometries, and active spaces. Its composition determines which variations in Hamiltonian structure contribute to the search objective. The resulting protocol can then be tested on Hamiltonians outside this ensemble. Each molecular instance has an electronic-structure Hamiltonian in a fixed spin-orbital active space,
\begin{equation}
H_{\rm el}
=
\sum_{pq} h_{pq} a_p^\dagger a_q
+
\frac{1}{2}
\sum_{pqrs} v_{pqrs} a_p^\dagger a_q^\dagger a_r a_s .
\label{eq:electronic_structure_hamiltonian}
\end{equation}
Fermion-to-qubit mappings, including Jordan--Wigner~\cite{jordan1928pauli} and Bravyi--Kitaev~\cite{bravyi2002fermionic}, express this operator as a weighted sum of Pauli strings,
\begin{equation}
H
=
\sum_{g=1}^{m}c_gP_g.
\label{eq:pf_pauli_hamiltonian}
\end{equation}
For the experiments reported here, we draw the training ensemble from HamLib~\cite{sawaya2024hamlib}, use its Jordan--Wigner representations, and account for the identity contribution analytically. The protocol constructs each circuit from the Hamiltonian structure and its fixed starting parameters. The nonuniform coefficients motivate synthesis tolerances that vary between rotations, although coefficient magnitude alone does not determine energy sensitivity.

Our tight baseline follows the procedure of Reiher \emph{et al.}~\cite{reiher2017reaction} and uses a nested second-order PF~\cite{trotter1959,strang1968}. For each training Hamiltonian, we numerically determine the repetitions needed to meet a reference Trotter-error target against its exact ground-state energy. Using this calibration, we minimize their cost model by allocating the chemical-accuracy budget among Trotter, rotation-synthesis, and QPE errors. This determines the final repetition count \(r\) and a uniform synthesis tolerance \(\epsilon_0\) shared by all rotations in the baseline circuit. Appendix~\ref{app:standard_pf_baseline} gives details on this construction.

\section{Results}
\label{sec:experiments}

We optimize one shared PF protocol on 115 HamLib molecular Hamiltonians spanning 4--16 qubits~\cite{sawaya2024hamlib} and a reaction-energy version on ten reactions. Five independent searches optimize the molecular protocol. Each candidate's rotations are compiled with CUDA-Q, and GPU Lanczos extracts its energies by applying the compiled circuit to state vectors without needing to construct the full unitary matrix. Molecular cases are evaluated on L4 and A100 GPUs per iteration, over 100 iterations for each seed.

\subsection{Energy estimates for molecular Hamiltonians}
\label{sec:experiments_hamlib}

The training set contains every Jordan--Wigner Hamiltonian with at most 16 qubits in the 663-instance HamLib electronic-structure dataset (Figure~\ref{fig:hamlib_molecular_dataset}). It covers 19 molecular species at different geometries and active-space sizes, with 2--14 electrons and 15--2913 listed Pauli terms, including the identity. All 115 Hamiltonians contribute to the search objective (cf. Equation \ref{eq:shared_protocol_objective}). The training results therefore describe performance on the same instances used to optimize the protocol. To check that the resulting protocol can still produce meaningful resource savings outside the training set, we test the unchanged ELOTE on the 76 Jordan--Wigner Hamiltonians with 18--22 qubits in the same dataset. The parameters of the baseline PF are predicted from a model fitted only to the training calibrations, and thus can give worse $T_{95}^{(0)}$ values on Hamiltonians from the test set. The test set contains new Hamiltonian instances, although some molecular species also appear in training with smaller active spaces. 

\begin{figure}
\centering
\includegraphics[width=\linewidth]{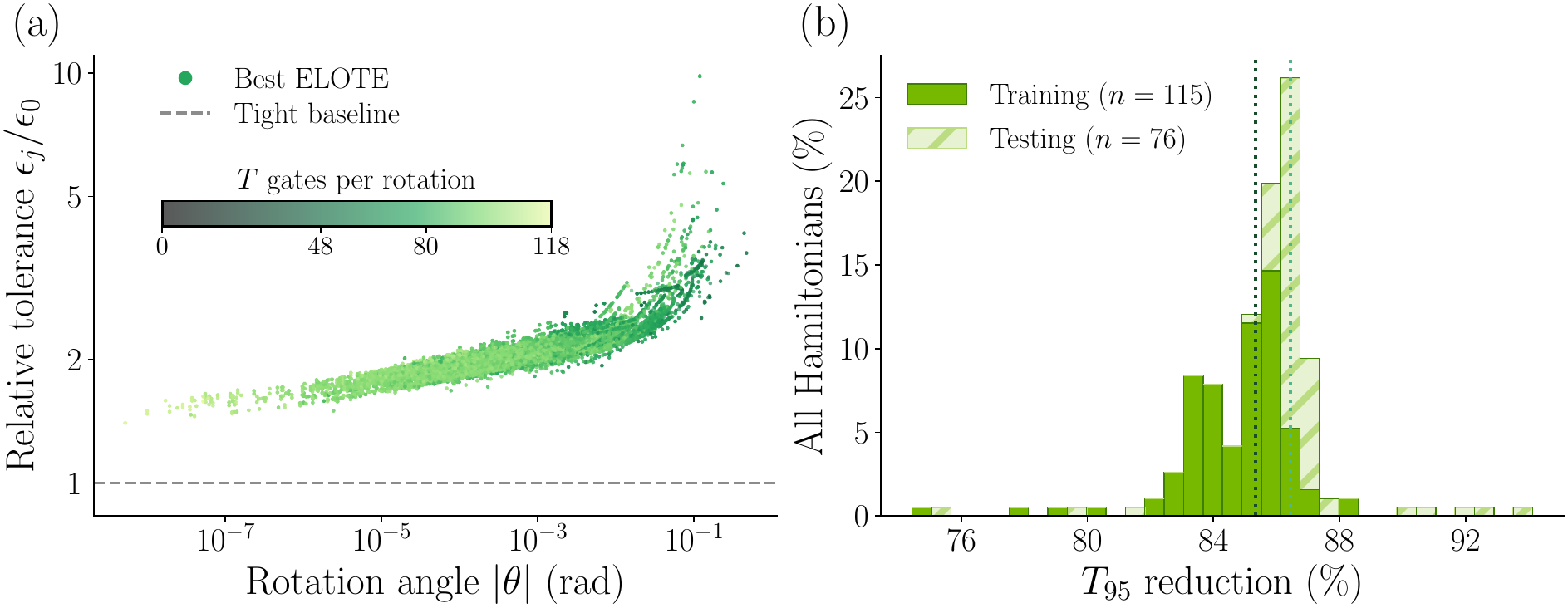}
\caption{Precision allocation and QPE-inclusive savings of ELOTE. (a) Relative synthesis tolerance $\epsilon_j/\epsilon_0$ versus rotation angle $|\theta|$ on the training set, colored by the compiled $\text{T}$ count per rotation. (b) Distribution of reductions in recorded $T_{95}$ cost estimates relative to the tight baseline, including QPE queries and repeated attempts. Dotted lines mark the training and testing medians.}
\label{fig:hamlib_pf_precision_reduction}
\end{figure}
Each search starts from the same tight baseline (Appendix~\ref{app:standard_pf_baseline}). The selected ELOTE applies the Hamiltonian terms one after another in a first-order formula, replacing the baseline's nested second-order splitting. It retains each Hamiltonian’s calibrated number of Trotter steps, combines contributions to the same Pauli string into a single term, and assigns each rotation a synthesis tolerance using an empirical formula discovered by the coding agent (Figure~\ref{fig:hamlib_pf_precision_reduction}a). On the 115 training Hamiltonians, the median reduction in QPE-inclusive $T_{95}$ is $85.35\%$ (Figure~\ref{fig:hamlib_pf_precision_reduction}b). For this selected PF, the geometric mean of baseline-relative $T_{95}$ across the 115 training Hamiltonians is $0.1518$, a reduction of $84.82\%$. Its mean and maximum absolute energy errors across the training set, before QPE readout, are $0.096$ and $0.356$~mHa (Figure~\ref{fig:hamlib_pf_search_trajectories}b). In Appendix~\ref{app:understanding_elote} we explain how ELOTE constructs its circuits and assigns tolerances, and also report statistics for each agentic loop.

For the 76 testing Hamiltonians on which the baseline can be constructed, the median reduction in the recorded $T_{95}$ cost estimate is $86.44\%$ (Figure~\ref{fig:hamlib_pf_precision_reduction}b). The baseline PF uses parameters fitted to the training Hamiltonians and can fail to meet chemical accuracy on the test set. ELOTE meets chemical accuracy in 76 of 76 cases, with a maximum error of $0.680$~mHa, whereas the baseline passes 75 of 76 cases. Its error for 18-qubit hydrogen fluoride is $1.833$~mHa, above the $1.594$~mHa limit (Figure~\ref{fig:hamlib_pf_search_trajectories}b). Among the 75 pairs in which both policies meet accuracy, ELOTE's geometric-mean baseline-relative $T_{95}$ is $0.1319$, a $86.81\%$ reduction.

Independent searches explore different parts of the protocol space and achieve different savings. The range between the best and worst searches widens overall as mutations accumulate (Figure~\ref{fig:hamlib_pf_search_trajectories}a). This variation suggests that more guided mutations could make the search more consistent.

\begin{figure}
\centering
\includegraphics[width=\linewidth]{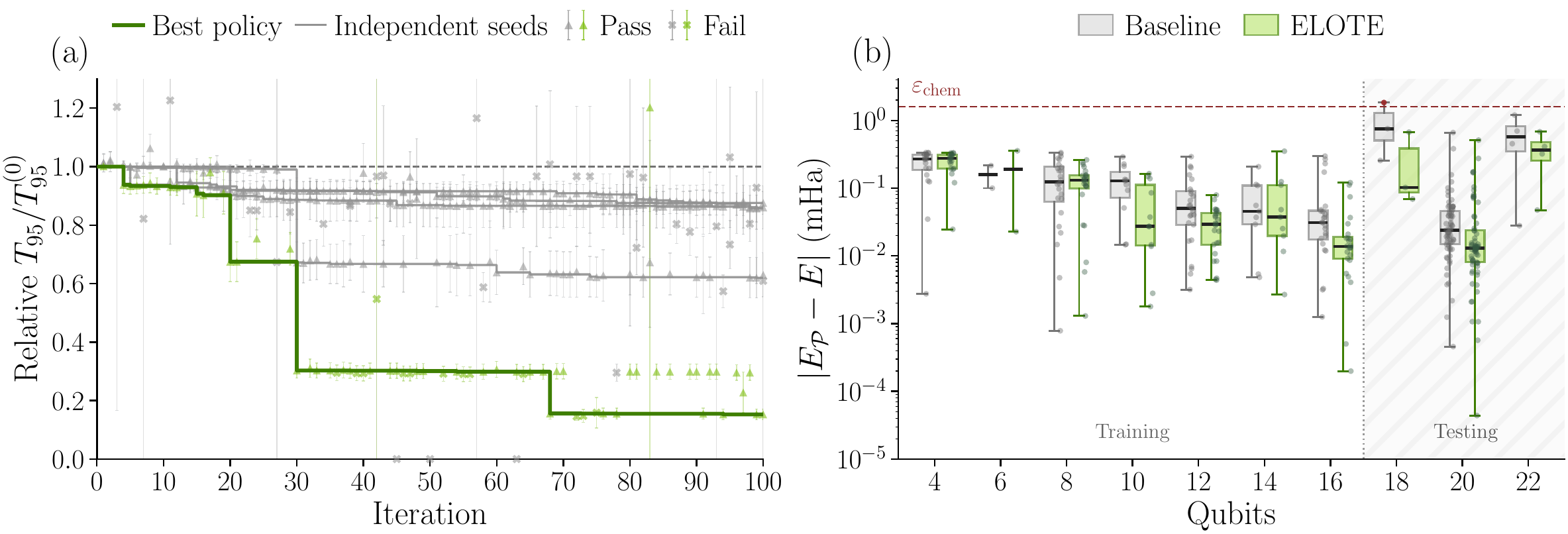}
\caption{Protocol optimization and energy accuracy. (a) Curves show mean relative $T_{95}$ over 115 Hamiltonians for five seeds. Green marks the leading seed. Markers show proposal means $\pm1$ SD across Hamiltonians. Triangles pass chemical accuracy for every Hamiltonian and crosses fail. Proposals with failed or inconclusive evaluations, or mean ratios above the plot's upper limit, are omitted. (b) Compiled energy errors for the baseline and ELOTE. Boxes span the middle 50\% and whiskers the full range.}
\label{fig:hamlib_pf_search_trajectories}
\end{figure}
\FloatBarrier

\subsection{Energy differences for chemical reactions}
\label{sec:experiments_reaction}

\begin{figure}
\centering
\includegraphics[width=0.613\linewidth]{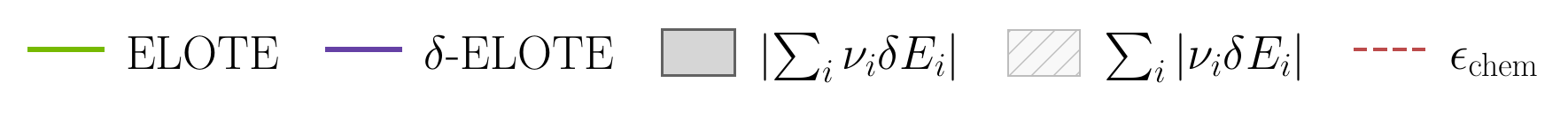}\\[-0.25em]
\sbox{\reactionkeybox}{%
\tcbox[colback=black!6,colframe=black!35,arc=3pt,
  boxrule=0.45pt,boxsep=0pt,left=4pt,right=4pt,top=5pt,bottom=5pt]{%
\footnotesize\begin{tabular}{@{}r@{\;}l@{}}
1. & $\mathrm{HCCH}\!\to\!\mathrm{H}_{2}\mathrm{CC}$ \\
2. & $\mathrm{HCN}\!\to\!\mathrm{HNC}$ \\
3. & $\mathrm{HCO}^{+}\!\to\!\mathrm{HOC}^{+}$ \\
4. & $\mathrm{CO}+\mathrm{H}_{2}\mathrm{O}\!\to\!\mathrm{CO}_{2}+\mathrm{H}_{2}$ \\
5. & $\mathrm{CO}+3\mathrm{H}_{2}\!\to\!\mathrm{CH}_{4}+\mathrm{H}_{2}\mathrm{O}$ \\
6. & $\mathrm{N}_{2}+3\mathrm{H}_{2}\!\to\!2\mathrm{NH}_{3}$ \\
7. & $\mathrm{H}_{2}+\mathrm{F}_{2}\!\to\!2\mathrm{HF}$ \\
8. & $\mathrm{H}_{2}+\mathrm{Cl}_{2}\!\to\!2\mathrm{HCl}$ \\
9. & $\mathrm{H}_{2}\mathrm{CO}\!\to\!\mathrm{HCOH}$ \\
10. & $\mathrm{HNO}\!\to\!\mathrm{NOH}$ \\
\end{tabular}
}}
\begin{minipage}[c]{\dimexpr(\linewidth-\wd\reactionkeybox-0.8em)/2\relax}
\centering
\includegraphics[width=\linewidth]{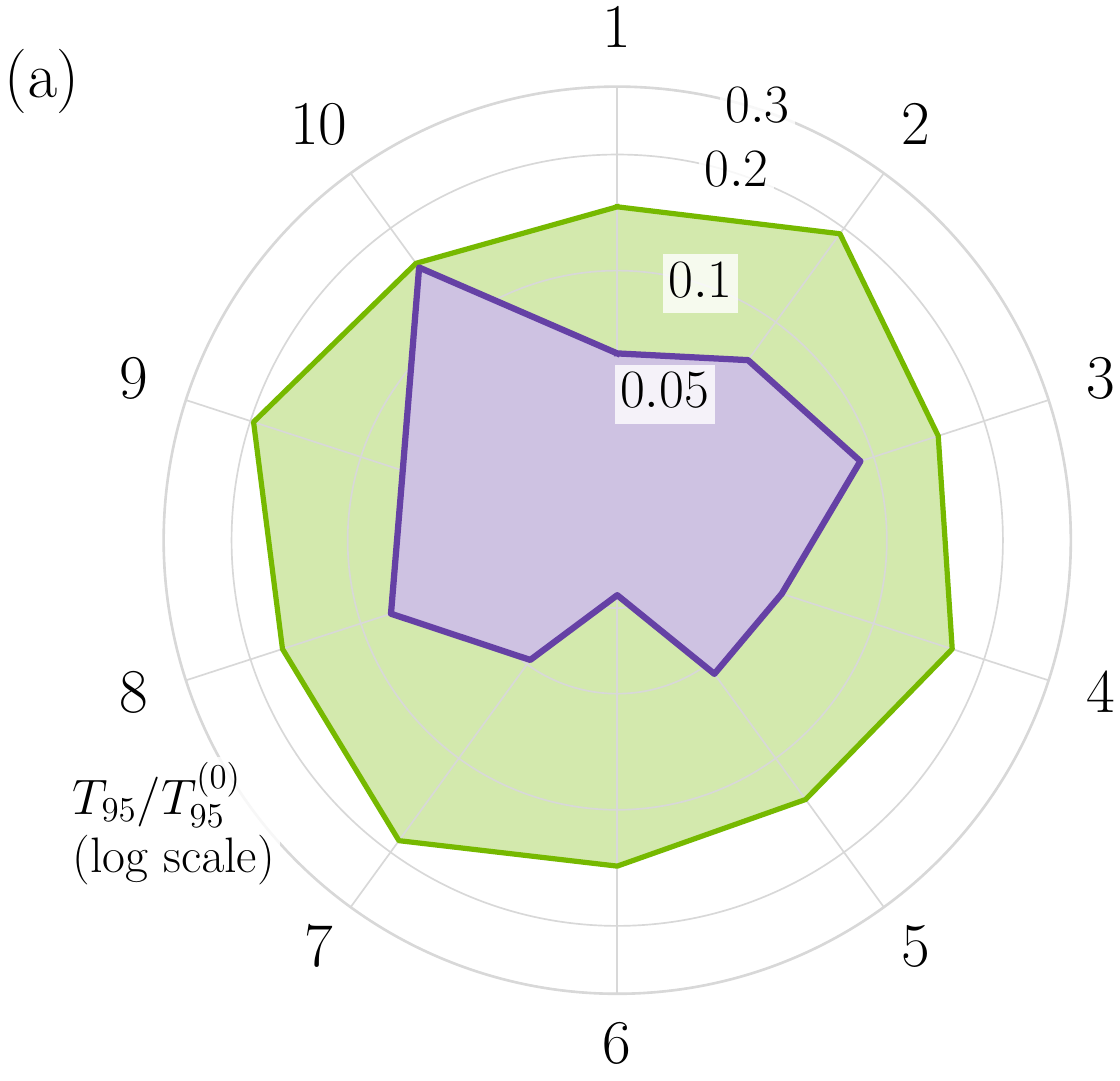}
\end{minipage}\hfill
\begin{minipage}[c]{\dimexpr(\linewidth-\wd\reactionkeybox-0.8em)/2\relax}
\centering
\includegraphics[width=\linewidth]{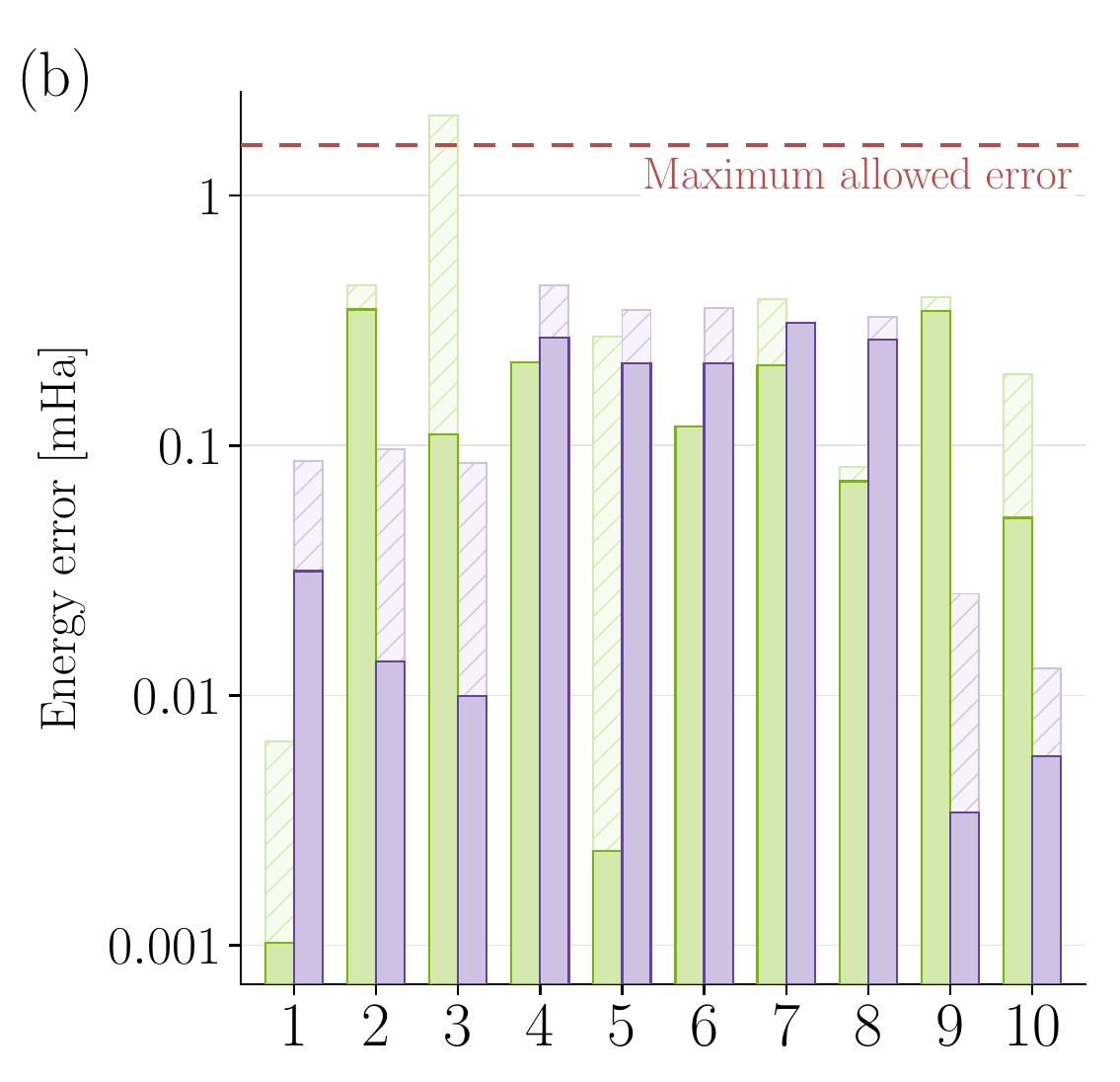}
\end{minipage}\hfill
\begin{minipage}[c]{\wd\reactionkeybox}
\usebox{\reactionkeybox}
\end{minipage}
\caption{Shared reaction-protocol optimization across ten reactions, numbered clockwise from the top in (a) and left to right in (b). (a) QPE-inclusive $T_{95}$ relative to each reaction's tight-baseline cost, for ELOTE and the selected $\delta$-ELOTE. The tight-baseline ratio of one lies outside the displayed range. (b) Compiled reaction-energy errors for the same policies, with paired green and purple bars for each reaction. Solid bar heights show $|\sum_i\nu_i\delta E_i|$; the tops of the faint hatched extensions show $\sum_i|\nu_i\delta E_i|$, the bound without cancellation. The red dashed line marks the chemical-accuracy limit.}
\label{fig:elote_delta_reactions}
\end{figure}

A reaction energy is a difference between molecular energies. Errors in those energies can cancel when combined with their stoichiometric signs~\cite{baysmidt2026nanographenes}. We therefore optimize the reaction-energy error directly and require chemical accuracy for every reaction. We consider five isomerization energies and five balanced reaction energies, shown in Figure~\ref{fig:elote_delta_reactions}. For stoichiometric coefficients $\nu_i$, the observable error is
\begin{equation}
    \delta \Delta E = \sum_i \nu_i \delta E_i,
\end{equation}
where $\delta E_i=E_{i,\mathcal P}-E_i$ is the signed compiled energy error for species $i$, with exact energy $E_i$ and compiled eigenphase energy $E_{i,\mathcal P}$. We require $|\delta\Delta E|<\varepsilon_{\rm chem}$ for every reaction.

We initialize five independent 50-mutation searches from an earlier molecular ELOTE protocol and optimize one shared reaction protocol, $\delta$-ELOTE. The best resulting protocol reduces the geometric-mean $T_{95}$ across the ten reactions by $58.90\%$ relative to the initial protocol while keeping every reaction-energy error within chemical accuracy (Figure~\ref{fig:elote_delta_reactions}). In panel (b), the difference between the solid bar heights and the tops of the hatched extensions shows cancellation between stoichiometrically weighted species errors, even where both quantities lie below chemical accuracy. For the selected $\delta$-ELOTE, both quantities are below the threshold for all ten reactions. The sum of absolute species errors exceeds chemical accuracy for HCO$^+\!\to\!\ $HOC$^+$ in the initial protocol, although its absolute reaction-energy error remains below the threshold. Appendix~\ref{app:delta_elote_policy} gives the repetition, ordering, and synthesis-tolerance rules for the optimized $\delta$-ELOTE protocol.

\FloatBarrier

\section{Conclusion}

In this paper, we introduced a coding-agent search over executable Hamiltonian-simulation protocols that jointly optimized PF structure and rotation-synthesis precision under a fixed chemical-accuracy constraint. Its cost included QPE queries and success amplification. Across 115 HamLib Hamiltonians spanning 4--16 qubits, the selected ELOTE achieved a median $85.35\%$ reduction in the QPE-based $T_{95}$ cost and a geometric-mean baseline-relative cost of $0.1518$, while satisfying chemical accuracy for every case. The agent discovered a shared rule that allocates rotation-synthesis precision term by term, using coefficient magnitude, Pauli support, and diagonal or off-diagonal character. The resulting savings reflected this joint change in structure and precision. With both the policy and baseline parameter fit frozen from training, ELOTE also met chemical accuracy on all 76 evaluated 18--22-qubit testing Hamiltonians.

For five isomerization energies and five balanced reaction energies, the selected shared $\delta$-ELOTE protocol reduced the geometric-mean $T_{95}$ across the ten reactions by $58.90\%$ relative to ELOTE, while every signed energy difference remained within chemical accuracy. This shows that a protocol learned on molecular energies can be refined for reaction-energy differences. All evaluations used ideal ground-state inputs. The larger-Hamiltonian tests assessed transfer to larger active spaces, with some molecular species shared between training and testing. These tests do not provide a universal error guarantee. Applying the search to larger systems would require certified error proxies, such as analytical bounds or numerical estimates~\cite{childs2021trottererror,blunt2025montecarloapproachbound}, in place of exact classical references.

Interesting future directions include applying a similar framework of agents in a loop to explore precision choices for Hamiltonian simulation methods, like quantum signal processing (QSP) or linear combination of unitaries (LCU). Promising results in this direction can shape how quantum hardware and middleware are designed, such as having $\text{T}$ states factories at different fidelities for different levels of precision in rotation synthesis~\cite{litinski2019magicstate}. Joint optimization with synthesis tolerances would require accounting for distillation noise, factory throughput, and the total QPE failure probability. Comparisons between general coding agents and specialized agents for quantum tasks~\cite{cao2026qcaleval} could establish whether domain adaptation improves autoresearch proposals. Similar loops could also use quantum computers to evaluate the proposed circuits. In such case, to iterate quickly, we would need efficient energy-certification methods to construct a reliable proxy for the target energy error, leading us to better solutions in shorter times.

\section*{Acknowledgements}

The authors acknowledge helpful discussions with Yuri Alexeev, Abdul Aldossary, Mohsen Bagherimehrab,  Ignacio Gustin, Prabhav Desai, Manav Ramprasad, and Zijian Zhang.

\section*{Code Availability}

The source code, agent instructions, and reproduction scripts are available at \url{https://github.com/bestquark/hsopt}.

\bibliographystyle{hsopt-unsrtnat}
\bibliography{references}

@article{suzuki1992,
  author = {Suzuki, Masuo},
  title = {General Theory of Higher-Order Decomposition of Exponential Operators and Symplectic Integrators},
  journal = {Physics Letters A},
  volume = {165},
  number = {5--6},
  pages = {387--395},
  year = {1992},
  doi = {10.1016/0375-9601(92)90335-J},
  url = {https://doi.org/10.1016/0375-9601(92)90335-J}
}

@article{trotter1959,
  author = {Trotter, H. F.},
  title = {On the Product of Semi-Groups of Operators},
  journal = {Proceedings of the American Mathematical Society},
  volume = {10},
  number = {4},
  pages = {545--551},
  year = {1959},
  doi = {10.1090/S0002-9939-1959-0108732-6},
  url = {https://doi.org/10.1090/S0002-9939-1959-0108732-6}
}

@article{strang1968,
  author = {Strang, Gilbert},
  title = {On the Construction and Comparison of Difference Schemes},
  journal = {SIAM Journal on Numerical Analysis},
  volume = {5},
  number = {3},
  pages = {506--517},
  year = {1968},
  doi = {10.1137/0705041},
  url = {https://doi.org/10.1137/0705041}
}

@article{suzuki1990,
  author = {Suzuki, Masuo},
  title = {Fractal Decomposition of Exponential Operators with Applications to Many-Body Theories and Monte Carlo Simulations},
  journal = {Physics Letters A},
  volume = {146},
  number = {6},
  pages = {319--323},
  year = {1990},
  doi = {10.1016/0375-9601(90)90962-N},
  url = {https://doi.org/10.1016/0375-9601(90)90962-N}
}

@book{schleich2025quantum,
  author = {Schleich, Philipp and Mantilla Calder{\'o}n, Luis and Sun, Chong and Bagherimehrab, Mohsen and Aldossary, Abdulrahman and Kottmann, Jakob S. and Aspuru-Guzik, Al{\'a}n},
  title = {Quantum Computing for Quantum Chemistry},
  publisher = {American Chemical Society},
  year = {2025},
  doi = {10.1021/acsinfocus.7e9012},
  isbn = {9780841295964},
  url = {https://doi.org/10.1021/acsinfocus.7e9012}
}

@article{bagherimehrab2024cpf,
  author = {Bagherimehrab, Mohsen and Mantilla Calderon, Luis and Berry, Dominic W. and Schleich, Philipp and Vakili, Mohammad Ghazi and Aldossary, Abdulrahman and Campos Gonzalez Angulo, Jorge A. and Gorgulla, Christoph and Aspuru-Guzik, Alan},
  title = {Faster Algorithmic Quantum and Classical Simulations by Corrected Product Formulas},
  journal = {PRX Quantum},
  volume = {7},
  number = {3},
  pages = {033018},
  year = {2026},
  eprint = {2409.08265},
  archiveprefix = {arXiv},
  primaryclass = {quant-ph},
  doi = {10.1103/phk5-slkl},
  url = {https://doi.org/10.1103/phk5-slkl}
}

@article{kliuchnikov2013,
  author = {Kliuchnikov, Vadym and Maslov, Dmitri and Mosca, Michele},
  title = {Asymptotically Optimal Approximation of Single Qubit Unitaries by Clifford and {T} Circuits Using a Constant Number of Ancillary Qubits},
  journal = {Physical Review Letters},
  volume = {110},
  number = {19},
  pages = {190502},
  year = {2013},
  doi = {10.1103/PhysRevLett.110.190502},
  url = {https://doi.org/10.1103/PhysRevLett.110.190502}
}

@article{rossselinger2016,
  author = {Ross, Neil J. and Selinger, Peter},
  title = {Optimal Ancilla-Free Clifford+{T} Approximation of $z$-Rotations},
  journal = {Quantum Information and Computation},
  volume = {16},
  number = {11--12},
  pages = {901--953},
  year = {2016},
  doi = {10.26421/QIC16.11-12-1},
  url = {https://doi.org/10.26421/QIC16.11-12-1}
}

@misc{matsumoto2008normalform,
  author = {Matsumoto, Ken and Amano, Kazuyuki},
  title = {Representation of Quantum Circuits with {Clifford} and $\pi/8$ Gates},
  year = {2008},
  eprint = {0806.3834},
  archiveprefix = {arXiv},
  primaryclass = {quant-ph},
  doi = {10.48550/arXiv.0806.3834},
  url = {https://doi.org/10.48550/arXiv.0806.3834}
}

@article{bothe2026smallangle,
  author = {Bothe, Marius and S{\"u}nderhauf, Christoph and Witham, Michael J. and Campbell, Earl T. and Blunt, Nick S.},
  title = {More Efficient Clifford+{T} Synthesis for Small-Angle Rotations and Application to Trotterization},
  journal = {arXiv preprint arXiv:2605.31544},
  year = {2026},
  eprint = {2605.31544},
  archiveprefix = {arXiv},
  primaryclass = {quant-ph},
  doi = {10.48550/arXiv.2605.31544},
  url = {https://doi.org/10.48550/arXiv.2605.31544}
}

@article{berry2015taylor,
  author = {Berry, Dominic W. and Childs, Andrew M. and Cleve, Richard and Kothari, Robin and Somma, Rolando D.},
  title = {Simulating Hamiltonian Dynamics with a Truncated Taylor Series},
  journal = {Physical Review Letters},
  volume = {114},
  number = {9},
  pages = {090502},
  year = {2015},
  eprint = {1412.4687},
  archiveprefix = {arXiv},
  primaryclass = {quant-ph},
  doi = {10.1103/PhysRevLett.114.090502},
  url = {https://doi.org/10.1103/PhysRevLett.114.090502}
}

@article{childs2021trottererror,
  author = {Childs, Andrew M. and Su, Yuan and Tran, Minh C. and Wiebe, Nathan and Zhu, Shuchen},
  title = {Theory of Trotter Error with Commutator Scaling},
  journal = {Physical Review X},
  volume = {11},
  number = {1},
  pages = {011020},
  year = {2021},
  eprint = {1912.08854},
  archiveprefix = {arXiv},
  primaryclass = {quant-ph},
  doi = {10.1103/PhysRevX.11.011020},
  url = {https://doi.org/10.1103/PhysRevX.11.011020}
}

@article{low2019qubitization,
  author = {Low, Guang Hao and Chuang, Isaac L.},
  title = {Hamiltonian Simulation by Qubitization},
  journal = {Quantum},
  volume = {3},
  pages = {163},
  year = {2019},
  eprint = {1610.06546},
  archiveprefix = {arXiv},
  primaryclass = {quant-ph},
  doi = {10.22331/q-2019-07-12-163},
  url = {https://doi.org/10.22331/q-2019-07-12-163}
}

@inproceedings{gilyen2019qsvt,
  author = {Gily{\'e}n, Andr{\'a}s and Su, Yuan and Low, Guang Hao and Wiebe, Nathan},
  title = {Quantum Singular Value Transformation and Beyond: Exponential Improvements for Quantum Matrix Arithmetics},
  booktitle = {Proceedings of the 51st Annual {ACM} {SIGACT} Symposium on Theory of Computing},
  pages = {193--204},
  year = {2019},
  eprint = {1806.01838},
  archiveprefix = {arXiv},
  primaryclass = {quant-ph},
  doi = {10.1145/3313276.3316366},
  url = {https://doi.org/10.1145/3313276.3316366}
}

@article{berry2019qubitizationchem,
  author = {Berry, Dominic W. and Gidney, Craig and Motta, Mario and McClean, Jarrod R. and Babbush, Ryan},
  title = {Qubitization of Arbitrary Basis Quantum Chemistry Leveraging Sparsity and Low Rank Factorization},
  journal = {Quantum},
  volume = {3},
  pages = {208},
  year = {2019},
  eprint = {1902.02134},
  archiveprefix = {arXiv},
  primaryclass = {quant-ph},
  doi = {10.22331/q-2019-12-02-208},
  url = {https://doi.org/10.22331/q-2019-12-02-208}
}

@article{babbush2018linear,
  author = {Babbush, Ryan and Gidney, Craig and Berry, Dominic W. and Wiebe, Nathan and McClean, Jarrod and Paler, Alexandru and Fowler, Austin and Neven, Hartmut},
  title = {Encoding Electronic Spectra in Quantum Circuits with Linear {T} Complexity},
  journal = {Physical Review X},
  volume = {8},
  number = {4},
  pages = {041015},
  year = {2018},
  eprint = {1805.03662},
  archiveprefix = {arXiv},
  primaryclass = {quant-ph},
  doi = {10.1103/PhysRevX.8.041015},
  url = {https://doi.org/10.1103/PhysRevX.8.041015}
}

@article{lee2021thc,
  author = {Lee, Joonho and Berry, Dominic W. and Gidney, Craig and Huggins, William J. and McClean, Jarrod R. and Wiebe, Nathan and Babbush, Ryan},
  title = {Even More Efficient Quantum Computations of Chemistry Through Tensor Hypercontraction},
  journal = {{PRX} Quantum},
  volume = {2},
  number = {3},
  pages = {030305},
  year = {2021},
  eprint = {2011.03494},
  archiveprefix = {arXiv},
  primaryclass = {quant-ph},
  doi = {10.1103/PRXQuantum.2.030305},
  url = {https://doi.org/10.1103/PRXQuantum.2.030305}
}

@article{highammary2022,
  author = {Higham, Nicholas J. and Mary, Th{\'e}o},
  title = {Mixed Precision Algorithms in Numerical Linear Algebra},
  journal = {Acta Numerica},
  volume = {31},
  pages = {347--414},
  year = {2022},
  doi = {10.1017/S0962492922000022},
  url = {https://doi.org/10.1017/S0962492922000022}
}

@inproceedings{micikevicius2018mixed,
  author = {Micikevicius, Paulius and Narang, Sharan and Alben, Jonah and Diamos, Gregory and Elsen, Erich and Garcia, David and Ginsburg, Boris and Houston, Michael and Kuchaiev, Oleksii and Venkatesh, Ganesh and Wu, Hao},
  title = {Mixed Precision Training},
  booktitle = {International Conference on Learning Representations},
  year = {2018},
  eprint = {1710.03740},
  archiveprefix = {arXiv},
  primaryclass = {cs.AI},
  doi = {10.48550/arXiv.1710.03740},
  url = {https://openreview.net/forum?id=r1gs9JgRZ}
}

@article{aspuruguzik2005simulated,
  author = {Aspuru-Guzik, Al{\'a}n and Dutoi, Anthony D. and Love, Peter J. and Head-Gordon, Martin},
  title = {Simulated Quantum Computation of Molecular Energies},
  journal = {Science},
  volume = {309},
  number = {5741},
  pages = {1704--1707},
  year = {2005},
  doi = {10.1126/science.1113479},
  url = {https://doi.org/10.1126/science.1113479}
}

@article{babbush2015chemical,
  author = {Babbush, Ryan and McClean, Jarrod and Wecker, Dave and Aspuru-Guzik, Al{\'a}n and Wiebe, Nathan},
  title = {Chemical Basis of {Trotter-Suzuki} Errors in Quantum Chemistry Simulation},
  journal = {Physical Review A},
  volume = {91},
  number = {2},
  pages = {022311},
  year = {2015},
  eprint = {1410.8159},
  archiveprefix = {arXiv},
  primaryclass = {quant-ph},
  doi = {10.1103/PhysRevA.91.022311},
  url = {https://doi.org/10.1103/PhysRevA.91.022311}
}

@article{lin2022heisenberg,
  author = {Lin, Lin and Tong, Yu},
  title = {Heisenberg-Limited Ground-State Energy Estimation for Early Fault-Tolerant Quantum Computers},
  journal = {{PRX} Quantum},
  volume = {3},
  number = {1},
  pages = {010318},
  year = {2022},
  doi = {10.1103/PRXQuantum.3.010318},
  url = {https://doi.org/10.1103/PRXQuantum.3.010318}
}

@article{reiher2017reaction,
  author = {Reiher, Markus and Wiebe, Nathan and Svore, Krysta M. and Wecker, Dave and Troyer, Matthias},
  title = {Elucidating Reaction Mechanisms on Quantum Computers},
  journal = {Proceedings of the National Academy of Sciences},
  volume = {114},
  number = {29},
  pages = {7555--7560},
  year = {2017},
  doi = {10.1073/pnas.1619152114},
  url = {https://doi.org/10.1073/pnas.1619152114}
}

@article{jones2012faster,
  author = {Jones, N. Cody and Whitfield, James D. and McMahon, Peter L. and Yung, Man-Hong and {Van Meter}, Rodney and Aspuru-Guzik, Al{\'a}n and Yamamoto, Yoshihisa},
  title = {Faster Quantum Chemistry Simulation on Fault-Tolerant Quantum Computers},
  journal = {New Journal of Physics},
  volume = {14},
  number = {11},
  pages = {115023},
  year = {2012},
  doi = {10.1088/1367-2630/14/11/115023},
  url = {https://doi.org/10.1088/1367-2630/14/11/115023}
}

@article{vonburg2021catalysis,
  author = {{von Burg}, Vera and Low, Guang Hao and H{\"a}ner, Thomas and Steiger, Damian S. and Reiher, Markus and Roetteler, Martin and Troyer, Matthias},
  title = {Quantum Computing Enhanced Computational Catalysis},
  journal = {Physical Review Research},
  volume = {3},
  number = {3},
  pages = {033055},
  year = {2021},
  doi = {10.1103/PhysRevResearch.3.033055},
  url = {https://doi.org/10.1103/PhysRevResearch.3.033055}
}

@article{goings2022cytochrome,
  author = {Goings, Joshua J. and White, Alec and Lee, Joonho and Tautermann, Christofer S. and Degroote, Matthias and Gidney, Craig and Shiozaki, Toru and Babbush, Ryan and Rubin, Nicholas C.},
  title = {Reliably Assessing the Electronic Structure of Cytochrome {P450} on Today's Classical Computers and Tomorrow's Quantum Computers},
  journal = {Proceedings of the National Academy of Sciences},
  volume = {119},
  number = {38},
  pages = {e2203533119},
  year = {2022},
  doi = {10.1073/pnas.2203533119},
  url = {https://doi.org/10.1073/pnas.2203533119}
}

@article{bellonzi2024catalyst,
  author = {Bellonzi, Nicole and Kunitsa, Alexander and Cantin, Joshua T. and Campos-Gonzalez-Angulo, Jorge A. and Radin, Maxwell D. and Zhou, Yanbing and Johnson, Peter D. and Mart{\'i}nez-Mart{\'i}nez, Luis A. and Jangrouei, Mohammad Reza and Brahmachari, Aritra Sankar and Wang, Linjun and Patel, Smik and Kodrycka, Monika and Loaiza, Ignacio and Lang, Robert A. and Aspuru-Guzik, Al{\'a}n and Izmaylov, Artur F. and Romero Fontalvo, Jhonathan and Cao, Yudong},
  title = {Feasibility of accelerating homogeneous catalyst discovery with fault-tolerant quantum computers},
  journal = {arXiv preprint arXiv:2406.06335},
  year = {2024},
  eprint = {2406.06335},
  archiveprefix = {arXiv},
  primaryclass = {quant-ph},
  doi = {10.48550/arXiv.2406.06335},
  url = {https://doi.org/10.48550/arXiv.2406.06335}
}

@article{sawaya2024hamlib,
  author = {Sawaya, Nicolas P. D. and Marti-Dafcik, Daniel and Ho, Yang and Tabor, Daniel P. and Bernal Neira, David E. and Magann, Alicia B. and Premaratne, Shavindra and Dubey, Pradeep and Matsuura, Anne and Bishop, Nathan and de Jong, Wibe A. and Benjamin, Simon and Parekh, Ojas and Tubman, Norm and Klymko, Katherine and Camps, Daan},
  title = {{HamLib}: A Library of Hamiltonians for Benchmarking Quantum Algorithms and Hardware},
  journal = {Quantum},
  volume = {8},
  pages = {1559},
  year = {2024},
  doi = {10.22331/q-2024-12-11-1559},
  url = {https://doi.org/10.22331/q-2024-12-11-1559}
}

@article{simon2025resource,
  author = {Simon, William A. and Love, Peter J.},
  title = {Quantum Advantage in Resource Estimation},
  journal = {arXiv preprint arXiv:2512.02131},
  year = {2025},
  eprint = {2512.02131},
  archiveprefix = {arXiv},
  doi = {10.48550/arXiv.2512.02131},
  url = {https://doi.org/10.48550/arXiv.2512.02131}
}

@article{cain2026shor,
  author = {Cain, Madelyn and Xu, Qian and King, Robbie and Picard, Lewis R. B. and Levine, Harry and Endres, Manuel and Preskill, John and Huang, Hsin-Yuan and Bluvstein, Dolev},
  title = {Shor's Algorithm Is Possible with as Few as 10,000 Reconfigurable Atomic Qubits},
  journal = {arXiv preprint arXiv:2603.28627},
  year = {2026},
  eprint = {2603.28627},
  archiveprefix = {arXiv},
  doi = {10.48550/arXiv.2603.28627},
  url = {https://doi.org/10.48550/arXiv.2603.28627}
}

@article{bravyi2002fermionic,
  author = {Bravyi, Sergey and Kitaev, Alexei},
  title = {Fermionic Quantum Computation},
  journal = {Annals of Physics},
  volume = {298},
  number = {1},
  pages = {210--226},
  year = {2002},
  eprint = {quant-ph/0003137},
  archiveprefix = {arXiv},
  primaryclass = {quant-ph},
  doi = {10.1006/aphy.2002.6254},
  url = {https://doi.org/10.1006/aphy.2002.6254}
}

@article{jordan1928pauli,
  author = {Jordan, Pascual and Wigner, Eugene},
  title = {{\"U}ber das Paulische {\"A}quivalenzverbot},
  journal = {Zeitschrift f{\"u}r Physik},
  volume = {47},
  number = {9--10},
  pages = {631--651},
  year = {1928},
  doi = {10.1007/BF01331938},
  url = {https://doi.org/10.1007/BF01331938}
}

@article{fawzi2022alphatensor,
  author = {Fawzi, Alhussein and Balog, Matej and Huang, Aja and Hubert, Thomas and Romera-Paredes, Bernardino and Barekatain, Mohammadamin and Novikov, Alexander and Ruiz, Francisco J. R. and Schrittwieser, Julian and Swirszcz, Grzegorz and Silver, David and Hassabis, Demis and Kohli, Pushmeet},
  title = {Discovering faster matrix multiplication algorithms with reinforcement learning},
  journal = {Nature},
  volume = {610},
  pages = {47--53},
  year = {2022},
  doi = {10.1038/s41586-022-05172-4},
  url = {https://doi.org/10.1038/s41586-022-05172-4}
}

@article{mankowitz2023alphadev,
  author = {Mankowitz, Daniel J. and Michi, Andrea and Zhernov, Anton and Gelmi, Marco and Selvi, Marco and Paduraru, Cosmin and Leurent, Edouard and Iqbal, Shariq and Lespiau, Jean-Baptiste and Ahern, Alex and Koppe, Thomas and Millikin, Kevin and Gaffney, Stephen and Elster, Sophie and Broshear, Jackson and Gamble, Chris and Milan, Kieran and Tung, Robert and Hwang, Minjae and Cemgil, Taylan and Barekatain, Mohammadamin and Li, Yujia and Mandhane, Amol and Hubert, Thomas and Schrittwieser, Julian and Hassabis, Demis and Kohli, Pushmeet and Riedmiller, Martin and Vinyals, Oriol and Silver, David},
  title = {Faster sorting algorithms discovered using deep reinforcement learning},
  journal = {Nature},
  volume = {618},
  pages = {257--263},
  year = {2023},
  doi = {10.1038/s41586-023-06004-9},
  url = {https://doi.org/10.1038/s41586-023-06004-9}
}

@article{romeraparedes2024funsearch,
  author = {Romera-Paredes, Bernardino and Barekatain, Mohammadamin and Novikov, Alexander and Balog, Matej and Kumar, M. Pawan and Dupont, Emilien and Ruiz, Francisco J. R. and Ellenberg, Jordan S. and Wang, Pengming and Fawzi, Omar and Holland, George and Kohli, Pushmeet and Fawzi, Alhussein},
  title = {Mathematical discoveries from program search with large language models},
  journal = {Nature},
  volume = {625},
  pages = {468--475},
  year = {2024},
  doi = {10.1038/s41586-023-06924-6},
  url = {https://doi.org/10.1038/s41586-023-06924-6}
}

@article{novikov2025alphaevolve,
  author = {Novikov, Alexander and V{\~u}, Ng{\^a}n and Eisenberger, Marvin and Dupont, Emilien and Huang, Po-Sen and Wagner, Adam Zsolt and Shirobokov, Sergey and Kozlovskii, Borislav and Ruiz, Francisco J. R. and Mehrabian, Abbas and Kumar, M. Pawan and See, Abigail and Chaudhuri, Swarat and Holland, George and Davies, Alex and Nowozin, Sebastian and Kohli, Pushmeet and Balog, Matej},
  title = {{AlphaEvolve}: A coding agent for scientific and algorithmic discovery},
  journal = {arXiv preprint arXiv:2506.13131},
  year = {2025},
  eprint = {2506.13131},
  archiveprefix = {arXiv},
  primaryclass = {cs.AI},
  doi = {10.48550/arXiv.2506.13131},
  url = {https://doi.org/10.48550/arXiv.2506.13131}
}

@article{zhang2025moleculargeometry,
  author = {Zhang, C. and Corti{\~n}as, R. G. and Karamlou, A. H. and Noll, N. and Provazza, J. and Bausch, J. and Shirobokov, S. and White, A. and Claassen, M. and Kang, S. H. and Senior, A. W. and Toma{\v{s}}ev, N. and Gross, J. and Lee, K. and Schuster, T. and Huggins, W. J. and others},
  title = {Quantum computation of molecular geometry via many-body nuclear spin echoes},
  journal = {arXiv preprint arXiv:2510.19550},
  year = {2025},
  eprint = {2510.19550},
  archiveprefix = {arXiv},
  primaryclass = {quant-ph},
  doi = {10.48550/arXiv.2510.19550},
  url = {https://doi.org/10.48550/arXiv.2510.19550}
}

@article{mantilla2026autoresearch,
  author = {Mantilla Calder{\'o}n, Luis and Gonthier, J{\'e}r{\^o}me F. and Gustin, Ignacio and Bernales, Varinia and Aspuru-Guzik, Al{\'a}n},
  title = {Optimizing ground state preparation protocols with autoresearch},
  journal = {arXiv preprint arXiv:2604.25610},
  year = {2026},
  eprint = {2604.25610},
  archiveprefix = {arXiv},
  primaryclass = {quant-ph},
  doi = {10.48550/arXiv.2604.25610},
  url = {https://doi.org/10.48550/arXiv.2604.25610}
}

@article{finger2026automated,
  author = {Finger, Fabian and Rapp, Frederic and Kalidindi, Pranav and He, Kerry and Yin, Kante and Koziell-Pipe, Alexander and Manrique, David Zsolt and Greene-Diniz, Gabriel and Clark, Stephen and Fawzi, Hamza and Romera-Paredes, Bernardino and Fawzi, Alhussein and Meichanetzidis, Konstantinos},
  title = {Automated near-term quantum algorithm discovery for molecular ground states},
  journal = {arXiv preprint arXiv:2603.26359},
  year = {2026},
  eprint = {2603.26359},
  archiveprefix = {arXiv},
  primaryclass = {quant-ph},
  doi = {10.48550/arXiv.2603.26359},
  url = {https://doi.org/10.48550/arXiv.2603.26359}
}

@misc{openevolve2025,
  author = {Sharma, Asankhaya},
  title = {{OpenEvolve}: Open-source implementation of an evolutionary coding agent},
  year = {2025},
  url = {https://github.com/algorithmicsuperintelligence/openevolve},
  urldate = {2026-07-21},
  note = {Algorithmic Superintelligence Inc.}
}

@inproceedings{ye2024reevo,
  author = {Berto, Federico and Cao, Zhiguang and Hua, Chuanbo and Kim, Haeyeon and Park, Jinkyoo and Song, Guojie and Wang, Jiarui and Ye, Haoran},
  title = {{ReEvo}: Large Language Models as Hyper-Heuristics with Reflective Evolution},
  booktitle = {Advances in Neural Information Processing Systems 37},
  pages = {43571--43608},
  year = {2024},
  eprint = {2402.01145},
  archiveprefix = {arXiv},
  primaryclass = {cs.AI},
  doi = {10.52202/079017-1381},
  url = {https://doi.org/10.52202/079017-1381}
}

@article{liu2024eoh,
  author = {Liu, Fei and Tong, Xialiang and Yuan, Mingxuan and Lin, Xi and Luo, Fu and Wang, Zhenkun and Lu, Zhichao and Zhang, Qingfu},
  title = {Evolution of Heuristics: Towards Efficient Automatic Algorithm Design Using Large Language Model},
  journal = {arXiv preprint arXiv:2401.02051},
  year = {2024},
  eprint = {2401.02051},
  archiveprefix = {arXiv},
  primaryclass = {cs.AI},
  doi = {10.48550/arXiv.2401.02051},
  note = {ICML 2024},
  url = {https://doi.org/10.48550/arXiv.2401.02051}
}

@inproceedings{chen2023evoprompting,
  author = {Chen, Angelica and Dohan, David M. and So, David R.},
  title = {{EvoPrompting}: Language Models for Code-Level Neural Architecture Search},
  booktitle = {Advances in Neural Information Processing Systems 36},
  pages = {7787--7817},
  year = {2023},
  eprint = {2302.14838},
  archiveprefix = {arXiv},
  primaryclass = {cs.CL},
  doi = {10.52202/075280-0342},
  url = {https://doi.org/10.52202/075280-0342}
}

@article{lange2025shinkaevolve,
  author = {Lange, Robert Tjarko and Imajuku, Yuki and Cetin, Edoardo},
  title = {{ShinkaEvolve}: Towards Open-Ended and Sample-Efficient Program Evolution},
  journal = {arXiv preprint arXiv:2509.19349},
  year = {2025},
  eprint = {2509.19349},
  archiveprefix = {arXiv},
  primaryclass = {cs.AI},
  doi = {10.48550/arXiv.2509.19349},
  url = {https://doi.org/10.48550/arXiv.2509.19349}
}

@article{yuksekgonul2026tttdiscover,
  author = {Y{\"u}ksekg{\"o}n{\"u}l, Mert and Koceja, Daniel and Li, Xinhao and Bianchi, Federico and McCaleb, Jed and Wang, Xiaolong and Kautz, Jan and Choi, Yejin and Zou, James and Guestrin, Carlos and Sun, Yu},
  title = {Learning to Discover at Test Time},
  journal = {arXiv preprint arXiv:2601.16175},
  year = {2026},
  eprint = {2601.16175},
  archiveprefix = {arXiv},
  primaryclass = {cs.LG},
  doi = {10.48550/arXiv.2601.16175},
  url = {https://doi.org/10.48550/arXiv.2601.16175}
}

@article{yu2025satlution,
  author = {Yu, Cunxi and Liang, Rongjian and Ho, Chia-Tung and Ren, Haoxing},
  title = {Autonomous Code Evolution Meets {NP}-Completeness},
  journal = {arXiv preprint arXiv:2509.07367},
  year = {2025},
  eprint = {2509.07367},
  archiveprefix = {arXiv},
  primaryclass = {cs.AI},
  doi = {10.48550/arXiv.2509.07367},
  url = {https://doi.org/10.48550/arXiv.2509.07367}
}

@inproceedings{zhang2025dgm,
  author = {Zhang, Jenny and Hu, Shengran and Lu, Cong and Lange, Robert and Clune, Jeff},
  title = {Darwin {G}{\"o}del Machine: Open-Ended Evolution of Self-Improving Agents},
  booktitle = {International Conference on Learning Representations},
  year = {2026},
  eprint = {2505.22954},
  archiveprefix = {arXiv},
  primaryclass = {cs.AI},
  doi = {10.48550/arXiv.2505.22954},
  url = {https://openreview.net/forum?id=pUpzQZTvGY}
}

@article{agrawal2025gepa,
  author = {Agrawal, Lakshya A. and Tan, Shangyin and Soylu, Dilara and Ziems, Noah and Khare, Rishi and Opsahl-Ong, Krista and Singhvi, Arnav and Shandilya, Herumb and Ryan, Michael J. and Jiang, Meng and Potts, Christopher and Sen, Koushik and Dimakis, Alexandros G. and Stoica, Ion and Klein, Dan and Zaharia, Matei and Khattab, Omar},
  title = {{GEPA}: Reflective Prompt Evolution Can Outperform Reinforcement Learning},
  journal = {arXiv preprint arXiv:2507.19457},
  year = {2025},
  eprint = {2507.19457},
  archiveprefix = {arXiv},
  primaryclass = {cs.CL},
  doi = {10.48550/arXiv.2507.19457},
  note = {Accepted to ICLR 2026 (Oral)},
  url = {https://doi.org/10.48550/arXiv.2507.19457}
}

@article{liu2026evox,
  author = {Liu, Shu and Agarwal, Shubham and Maheswaran, Monishwaran and Cemri, Mert and Li, Zhifei and Mang, Qiuyang and Naren, Ashwin and Boneh, Ethan and Cheng, Audrey and Pan, Melissa Z. and Du, Alexander and Keutzer, Kurt and Cheung, Alvin and Dimakis, Alexandros G. and Sen, Koushik and Zaharia, Matei and Stoica, Ion},
  title = {{EvoX}: Meta-Evolution for Automated Discovery},
  journal = {arXiv preprint arXiv:2602.23413},
  year = {2026},
  eprint = {2602.23413},
  archiveprefix = {arXiv},
  primaryclass = {cs.LG},
  doi = {10.48550/arXiv.2602.23413},
  url = {https://doi.org/10.48550/arXiv.2602.23413}
}

@software{opencode_ai,
  author  = {{Anomaly}},
  title   = {{OpenCode}: The Open Source {AI} Coding Agent},
  year    = {2026},
  version = {2.0.26},
  url     = {https://github.com/anomalyco/opencode}
}

@software{nvidia_openshell_2026,
  author  = {{NVIDIA}},
  title   = {{OpenShell}: The Safe, Private Runtime for Autonomous {AI} Agents},
  year    = {2026},
  version = {0.1.3},
  url     = {https://github.com/NVIDIA/OpenShell}
}

@article{brower2025blackwell,
  author = {Brower, Cole and Rodriguez Bernabeu, Samuel and Hammond, Jeff and Gunnels, John and Xantheas, Sotiris S. and Ganahl, Martin and Menczer, Andor and Legeza, {\"O}rs},
  title = {Mixed-precision ab initio tensor network state methods adapted for {NVIDIA} {Blackwell} technology via emulated {FP64} arithmetic},
  journal = {Journal of Chemical Theory and Computation},
  volume = {22},
  number = {13},
  pages = {6275--6282},
  year = {2026},
  eprint = {2510.04795},
  archiveprefix = {arXiv},
  primaryclass = {physics.chem-ph},
  doi = {10.1021/acs.jctc.6c00203},
  url = {https://doi.org/10.1021/acs.jctc.6c00203}
}

@misc{nvidia2025nvfp4,
  author = {Alvarez, Eduardo and Almog, Omri and Chung, Eric and Layton, Simon and Stosic, Dusan and Krashinsky, Ronny and Aubrey, Kyle},
  title = {Introducing {NVFP4} for Efficient and Accurate Low-Precision Inference},
  year = {2025},
  month = jun,
  howpublished = {NVIDIA Technical Blog},
  url = {https://developer.nvidia.com/blog/introducing-nvfp4-for-efficient-and-accurate-low-precision-inference/},
  urldate = {2026-10-27}
}

@article{litinski2019magicstate,
  author = {Litinski, Daniel},
  title = {Magic State Distillation: Not as Costly as You Think},
  journal = {Quantum},
  volume = {3},
  pages = {205},
  year = {2019},
  doi = {10.22331/q-2019-12-02-205},
  url = {https://doi.org/10.22331/q-2019-12-02-205}
}

@article{zhang2026hyperagents,
  author = {Zhang, Jenny and Zhao, Bingchen and Yang, Wannan and Foerster, Jakob and Clune, Jeff and Jiang, Minqi and Devlin, Sam and Shavrina, Tatiana},
  title = {Hyperagents},
  journal = {arXiv preprint arXiv:2603.19461},
  year = {2026},
  eprint = {2603.19461},
  archiveprefix = {arXiv},
  primaryclass = {cs.AI},
  doi = {10.48550/arXiv.2603.19461},
  url = {https://doi.org/10.48550/arXiv.2603.19461}
}

@book{nielsen2010quantum,
  author = {Nielsen, Michael A. and Chuang, Isaac L.},
  title = {Quantum Computation and Quantum Information},
  publisher = {Cambridge University Press},
  year = {2010},
  doi = {10.1017/cbo9780511976667},
  isbn = {9781107002173},
  url = {https://doi.org/10.1017/cbo9780511976667}
}

@misc{nvidia_cudaq_rotation_synthesis,
  author = {{NVIDIA}},
  title = {{CUDA-Q}: Rotation Synthesis ({Clifford}+{T})},
  year = {2026},
  url = {https://nvidia.github.io/cuda-quantum/latest/using/examples/rotation_synthesis.html},
  note = {Accessed October 11, 2026}
}

@article{gustin2025elagente,
  author = {Gustin, Ignacio and Mantilla Calder{\'o}n, Luis and P{\'e}rez-S{\'a}nchez, Juan B. and Crebolder, Chris and Gonthier, J{\'e}r{\^o}me F. and Ghazi Vakili, Mohammad and Nakamura, Yuma and Panicker, Karthik and Ramprasad, Manav and Yin, Aiwei and Zhang, Zijian and Zou, Yunheng and Bernales, Varinia and Aspuru-Guzik, Al{\'a}n},
  title = {El Agente Cu{\'a}ntico: Automating quantum simulations},
  journal = {Reports on Progress in Physics},
  volume = {89},
  number = {7},
  pages = {077602},
  year = {2026},
  eprint = {2512.18847},
  archiveprefix = {arXiv},
  primaryclass = {quant-ph},
  doi = {10.1088/1361-6633/ae8933},
  url = {https://doi.org/10.1088/1361-6633/ae8933}
}

@article{zou2025elagente,
  author = {Zou, Yunheng and Cheng, Austin H. and Aldossary, Abdulrahman and Bai, Jiaru and Leong, Shi Xuan and Campos-Gonzalez-Angulo, Jorge Arturo and Choi, Changhyeok and Ser, Cher Tian and Tom, Gary and Wang, Andrew and Zhang, Zijian and Yakavets, Ilya and Hao, Han and Crebolder, Chris and Bernales, Varinia and Aspuru-Guzik, Al{\'a}n},
  title = {{El Agente}: An autonomous agent for quantum chemistry},
  journal = {Matter},
  volume = {8},
  number = {7},
  pages = {102263},
  year = {2025},
  doi = {10.1016/j.matt.2025.102263},
  url = {https://doi.org/10.1016/j.matt.2025.102263}
}

@article{pham2026chemgraph,
  author = {Pham, Thang D. and Tanikanti, Aditya and Ke{\c c}eli, Murat},
  title = {{ChemGraph} as an agentic framework for computational chemistry workflows},
  journal = {Communications Chemistry},
  volume = {9},
  number = {1},
  pages = {33},
  year = {2026},
  doi = {10.1038/s42004-025-01776-9},
  url = {https://doi.org/10.1038/s42004-025-01776-9}
}

@inproceedings{fu2026qagent,
  author = {Fu, Zhenxiao and Jiang, Lei and Xu, Yilun and Huang, Gang and Chen, Fan},
  title = {{QAgent}: An {LLM}-based Multi-Agent System for Autonomous {OpenQASM} Programming},
  booktitle = {2026 IEEE International Conference on Quantum Computing and Engineering (QCE)},
  year = {2026},
  url = {https://doi.org/10.48550/arXiv.2508.20134},
  doi = {10.48550/arXiv.2508.20134}
}

@misc{blunt2025montecarloapproachbound,
  author = {Blunt, Nick S. and Ivanov, Aleksei V. and Bay-Smidt, Andreas Juul},
  title = {A Monte Carlo approach to bound Trotter error},
  year = {2025},
  eprint = {2510.11621},
  archiveprefix = {arXiv},
  primaryclass = {quant-ph},
  doi = {10.48550/arXiv.2510.11621},
  url = {https://doi.org/10.48550/arXiv.2510.11621}
}

@article{cao2026qcaleval,
  author = {Cao, Shuxiang and Zhang, Zijian and Agarwal, Abhishek and Bratrud, Grace and Beysengulov, Niyaz R. and Cole, Daniel C. and G{\'o}mez Frieiro, Alejandro and Glen, Elena O. and Hsu, Hao and Huang, Gang and Jow, Raymond and Shaji, Greshma and Lubowe, Tom and Zhu, Ligeng and Mantilla Calder{\'o}n, Luis and Pancotti, Nicola and Pendleton, Joel and Severin, Brandon and Staub, Charles Etienne and Sussman, Sara and Veps{\"a}l{\"a}inen, Antti and Vora, Neel Rajeshbhai and Xu, Yilun and Bernales, Varinia and Bowring, Daniel and Kyoseva, Elica and Rungger, Ivan and Semeghini, Giulia and Stanwyck, Sam and Costa, Timothy and Aspuru-Guzik, Al{\'a}n and Svore, Krysta},
  title = {{QCalEval}: Benchmarking Vision-Language Models for Quantum Calibration Plot Understanding},
  journal = {arXiv preprint arXiv:2604.25884},
  year = {2026},
  eprint = {2604.25884},
  archiveprefix = {arXiv},
  primaryclass = {quant-ph},
  doi = {10.48550/arXiv.2604.25884},
  url = {https://doi.org/10.48550/arXiv.2604.25884}
}

@article{baysmidt2026nanographenes,
  author = {Bay-Smidt, Andreas Juul and Glaser, Nina and Fabian, Marcel D. and Campbell, Earl T. and Blunt, Nick S. and Solomon, Gemma C.},
  title = {Quantum simulation of nanographenes and Trotter error cancellation},
  journal = {arXiv preprint arXiv:2605.00745},
  year = {2026},
  eprint = {2605.00745},
  archiveprefix = {arXiv},
  primaryclass = {quant-ph},
  doi = {10.48550/arXiv.2605.00745},
  url = {https://doi.org/10.48550/arXiv.2605.00745}
}

@article{liu2026solpi,
  author = {Liu, Haozhe and Ye, Tian and Gao, Sensen and Cao, Qihang and Li, Yitong and Zhuge, Mingchen and Wang, Duomin and Zhang, Ruihua and Luo, Ping and Bian, Jiawang and Zhu, Lei and Zhu, Ligeng and Xie, Enze and Han, Song},
  title = {{SoL-Pi}: Recursively Scaling Auto-Research Loops for Efficient Agent Harness},
  journal = {arXiv preprint arXiv:2609.20519},
  year = {2026},
  eprint = {2609.20519},
  archiveprefix = {arXiv},
  primaryclass = {cs.AI},
  doi = {10.48550/arXiv.2609.20519},
  url = {https://doi.org/10.48550/arXiv.2609.20519}
}

@article{gluza2018fidelitywitnesses,
  author = {Gluza, M. and Kliesch, M. and Eisert, J. and Aolita, L.},
  title = {Fidelity Witnesses for Fermionic Quantum Simulations},
  journal = {Physical Review Letters},
  volume = {120},
  number = {19},
  pages = {190501},
  year = {2018},
  eprint = {1703.03152},
  archiveprefix = {arXiv},
  primaryclass = {quant-ph},
  doi = {10.1103/PhysRevLett.120.190501},
  url = {https://doi.org/10.1103/PhysRevLett.120.190501}
}

@article{rall2019paulipropagation,
  author = {Rall, Patrick and Liang, Daniel and Cook, Jeremy and Kretschmer, William},
  title = {Simulation of Qubit Quantum Circuits via {Pauli} Propagation},
  journal = {Physical Review A},
  volume = {99},
  number = {6},
  pages = {062337},
  year = {2019},
  eprint = {1901.09070},
  archiveprefix = {arXiv},
  primaryclass = {quant-ph},
  doi = {10.1103/PhysRevA.99.062337},
  url = {https://doi.org/10.1103/PhysRevA.99.062337}
}

@article{rudolph2026paulipropagation,
  author = {Rudolph, Manuel S. and Jones, Tyson and Teng, Yanting and Angrisani, Armando and Holmes, Zo{\"e}},
  title = {{Pauli} Propagation: A Computational Framework for Simulating Quantum Systems},
  journal = {PRX Quantum},
  volume = {7},
  number = {3},
  pages = {032001},
  year = {2026},
  eprint = {2505.21606},
  archiveprefix = {arXiv},
  primaryclass = {quant-ph},
  doi = {10.1103/6vd7-l9bn},
  url = {https://doi.org/10.1103/6vd7-l9bn}
}

@article{miller2025majoranapropagation,
  author = {Miller, Aaron and Favre, Joachim and Holmes, Zo{\"e} and Salehi, {\"O}zlem and Chakraborty, Rahul and Nyk{\"a}nen, Anton and Zimbor{\'a}s, Zolt{\'a}n and Glos, Adam and Garc{\'i}a-P{\'e}rez, Guillermo},
  title = {Simulation of Fermionic Circuits Using {Majorana} Propagation},
  journal = {arXiv preprint arXiv:2503.18939},
  year = {2025},
  eprint = {2503.18939},
  archiveprefix = {arXiv},
  primaryclass = {quant-ph},
  doi = {10.48550/arXiv.2503.18939},
  url = {https://doi.org/10.48550/arXiv.2503.18939}
}

@article{mehendale2025trottererrors,
  author = {Mehendale, Shashank G. and Mart{\'i}nez-Mart{\'i}nez, Luis A. and Kamath, Prathami Divakar and Izmaylov, Artur F.},
  title = {Estimating {Trotter} Approximation Errors to Optimize Hamiltonian Partitioning for Lower Eigenvalue Errors},
  journal = {Digital Discovery},
  volume = {4},
  number = {12},
  pages = {3540--3551},
  year = {2025},
  eprint = {2312.13282},
  archiveprefix = {arXiv},
  primaryclass = {physics.chem-ph},
  doi = {10.1039/D5DD00185D},
  url = {https://doi.org/10.1039/D5DD00185D}
}

@article{maxwell2026practicaltrotter,
  author = {Maxwell, William and Casares, Pablo A. M. and Lang, Robert A. and Fomichev, Stepan and Arrazola, Juan Miguel and Jahangiri, Soran and Asadi, Ali and Nunez Meneses, Luis Alfredo and Germain, Thomas and Motlagh, Danial},
  title = {Practical Estimation of {Trotter} Error for Hamiltonian Simulation},
  journal = {arXiv preprint arXiv:2606.30738},
  year = {2026},
  eprint = {2606.30738},
  archiveprefix = {arXiv},
  primaryclass = {quant-ph},
  doi = {10.48550/arXiv.2606.30738},
  url = {https://doi.org/10.48550/arXiv.2606.30738}
}

@article{horodecki1996separability,
  author = {Horodecki, Micha{\l} and Horodecki, Pawe{\l} and Horodecki, Ryszard},
  title = {Separability of Mixed States: Necessary and Sufficient Conditions},
  journal = {Physics Letters A},
  volume = {223},
  number = {1--2},
  pages = {1--8},
  year = {1996},
  eprint = {quant-ph/9605038},
  archiveprefix = {arXiv},
  doi = {10.1016/S0375-9601(96)00706-2},
  url = {https://doi.org/10.1016/S0375-9601(96)00706-2}
}

@article{terhal2000witnesses,
  author = {Terhal, Barbara M.},
  title = {Bell Inequalities and the Separability Criterion},
  journal = {Physics Letters A},
  volume = {271},
  number = {5--6},
  pages = {319--326},
  year = {2000},
  eprint = {quant-ph/9911057},
  archiveprefix = {arXiv},
  doi = {10.1016/S0375-9601(00)00401-1},
  url = {https://doi.org/10.1016/S0375-9601(00)00401-1}
}

@article{gu2026quantumcircuits,
  author = {Gu, Andi and Kemeny, Alek H. and Yelin, Susanne F. and Lukin, Mikhail D.},
  title = {Discovering State-of-the-Art Quantum Circuits with Large Language Models},
  journal = {arXiv preprint arXiv:2610.22367},
  year = {2026},
  eprint = {2610.22367},
  archiveprefix = {arXiv},
  primaryclass = {quant-ph},
  doi = {10.48550/arXiv.2610.22367},
  url = {https://doi.org/10.48550/arXiv.2610.22367}
}

@unpublished{alpoge2026forcedeuler,
  author = {Alp{\"o}ge, Levent and Buckmaster, Tristan},
  title = {Blowup for the {Euler} Equations with Smooth Forcing},
  year = {2026},
  note = {Preprint},
  url = {https://cims.nyu.edu/~tristanb/euler.pdf},
  urldate = {2026-10-28}
}

@misc{openai2026navierstokes,
  author = {{OpenAI}},
  title = {On the {Navier--Stokes} Millennium Prize Problem},
  year = {2026},
  month = sep,
  note = {Research announcement},
  url = {https://openai.com/index/navier-stokes-solution/},
  urldate = {2026-10-28}
}

@misc{openai2026scientificunderstanding,
  author = {{OpenAI}},
  title = {Working with Academic Communities to Build Scientific Understanding},
  year = {2026},
  month = oct,
  note = {Research announcement},
  url = {https://openai.com/index/building-scientific-understanding-with-academic-communities/},
  urldate = {2026-10-28}
}

@article{georgiev2026mathematicaldiscovery,
  author = {Georgiev, Bogdan and G{\'o}mez-Serrano, Javier and Tao, Terence and Wagner, Adam Zsolt},
  title = {Mathematical Discovery and Exploration Can Be Done at Scale},
  journal = {Proceedings of the National Academy of Sciences},
  volume = {123},
  number = {40},
  pages = {e2536158123},
  year = {2026},
  doi = {10.1073/pnas.2536158123},
  url = {https://doi.org/10.1073/pnas.2536158123}
}

@techreport{nvidia2026nemotron35,
  author      = {{NVIDIA}},
  title       = {{Nemotron 3.5 Super and Lightning}: Open, Efficient Mixture-of-Experts Hybrid Mamba-Transformer Models for Agentic {AI}},
  institution = {NVIDIA},
  year        = {2026},
  month       = oct,
  url         = {https://research.nvidia.com/labs/nemotron/Nemotron-3.5-Super-and-Lightning/}
}

@article{king2026quantum,
  title={Quantum simulation with sum-of-squares spectral amplification},
  author={King, Robbie and Low, Guang Hao and Babbush, Ryan and Somma, Rolando D and Rubin, Nicholas C},
  journal={Physical Review Letters},
  volume={136},
  number={11},
  pages={110601},
  year={2026},
  publisher={APS}
}

@article{wang2025energycertification,
  author = {Wang, Guoming and Stilck Fran{\c{c}}a, Daniel and Rendon, Gumaro and Johnson, Peter D.},
  title = {Efficient Ground-State Energy Estimation and Certification on Early Fault-Tolerant Quantum Computers},
  journal = {Physical Review A},
  volume = {111},
  number = {1},
  pages = {012426},
  year = {2025},
  doi = {10.1103/PhysRevA.111.012426},
  url = {https://doi.org/10.1103/PhysRevA.111.012426}
}

\clearpage
\section*{\centering\LARGE Supplementary material}
\appendix
\setcounter{secnumdepth}{3}

\section{Benchmark and evaluation details}
\label{app:evaluator}

Here, we describe the benchmarks, energy and gate-cost calculations, and baseline PF construction.

\subsection{Datasets and energy evaluation}
\label{app:molecular_dataset}
\label{app:energy_evaluation}

\begin{figure}[!ht]
\centering
\includegraphics[width=0.94\linewidth]{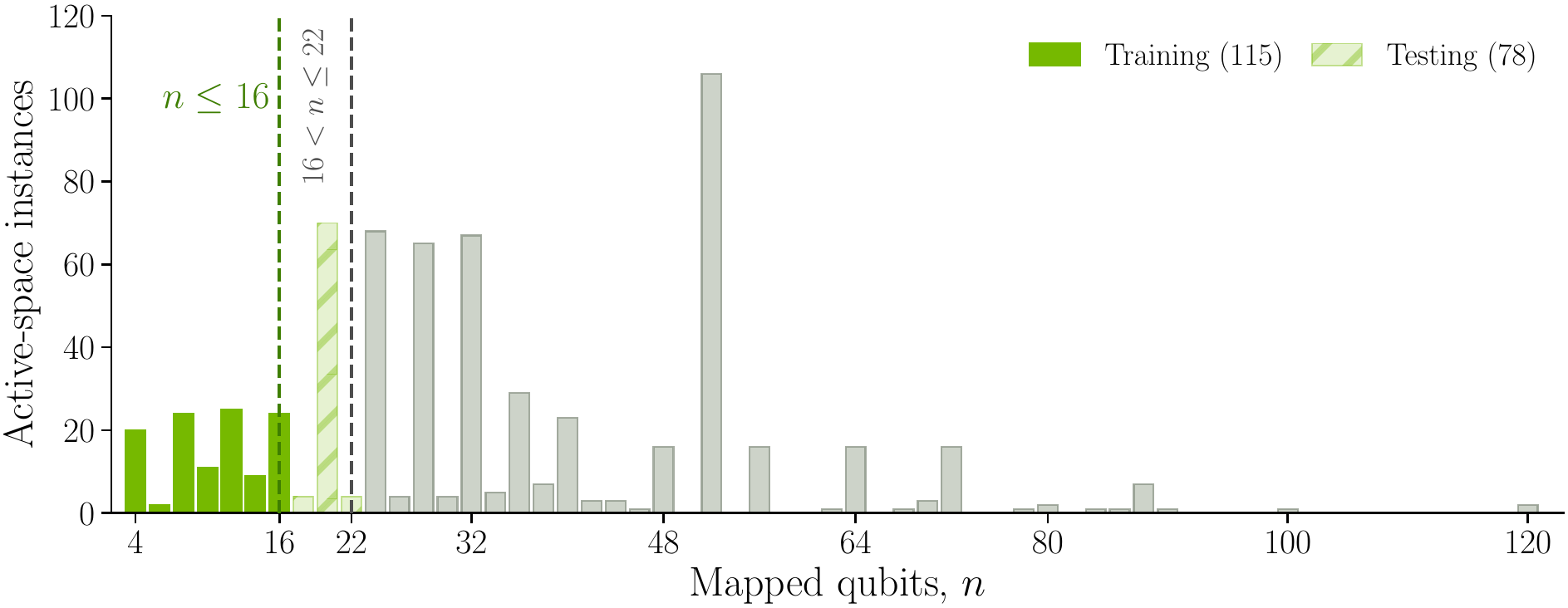}
\caption{All 663 HamLib electronic-structure instances by active-space size.
The training set contains 115 Hamiltonians with \(n\le16\), including 24 at
16 qubits. Of the 78 Hamiltonians in the testing range \(16<n\le22\), 76 were
evaluated in Figures~\ref{fig:hamlib_pf_precision_reduction}
and~\ref{fig:hamlib_pf_search_trajectories}.}
\label{fig:hamlib_molecular_dataset}
\end{figure}

Figure~\ref{fig:hamlib_molecular_dataset} shows the HamLib training and testing ranges~\cite{sawaya2024hamlib}. We excluded the 18- and 20-qubit NaLi Hamiltonians because the Pauli coefficients supplied by HamLib did not exactly reproduce the electron-number-conserving form used in our baseline PF reconstruction.

We constructed the reaction benchmark from 21 molecular species at fixed geometries in STO-3G, with fixed charges, spin multiplicities, active spaces, and stoichiometric coefficients. All five isomerizations and five balanced reactions in Figure~\ref{fig:elote_delta_reactions} served both optimization and evaluation. The Jordan--Wigner Hamiltonians use four qubits for H$_2$ and ten for every other species. We combined exact ground-state energies with stoichiometric weights to obtain reference electronic energy differences, excluding thermal corrections.

For each system, all proposals shared the Hamiltonian \(H=c_0I+\sum_{j=1}^{M}c_jP_j\), prescribed electron number, exact ground-state input \(H\ket{\psi_0}=E_0\ket{\psi_0}\), and evolution time \(t=1/\lambda\), with nonidentity \(P_j\) and \(\lambda=\sum_j|c_j|\). We added \(c_0\) analytically. The ideal input excludes state-preparation errors. CUDA-Q synthesized rotations in Matsumoto--Amano normal form~\cite{matsumoto2008normalform}.

Let \(\ket{v_j}\) be a normalized eigenvector of the compiled step, with
\(S_{\mathcal P}\ket{v_j}=e^{i\phi_j}\ket{v_j}\). For an input
\(\ket{\psi_0}\), the probability of measuring eigenphase \(\phi_j\) is
\(p_j=|\braket{v_j}{\psi_0}|^2\). We use matrix-free Lanczos to identify the most probable energy outcome, requiring its total probability to exceed \(1/2\). After \(r\) steps, its decoded energy is
\begin{equation}
\mathcal C_\star=\arg\max_{\mathcal C}\sum_{j\in\mathcal C}p_j,
\qquad
E_{\mathcal P}=c_0+\frac{\operatorname{Arg}(e^{ir\phi_\star})}{t}.
\label{eq:pf_energy_selection}
\end{equation}
Here \(\mathcal C_\star\) groups numerically indistinguishable energies, \(\phi_\star\) is a representative phase, and \(p_\star=\sum_{j\in\mathcal C_\star}p_j\). We require \(p_\star>1/2\), ensuring that no other outcome is more probable. We check how closely each selected vector satisfies the compiled-step eigenvalue equation and require independent Lanczos starts to agree in energy and probability within \(10^{-10}\,\mathrm{Ha}\) and \(10^{-9}\), respectively. We also require \(|r\phi_j|<\pi-10^{-6}\) for the selected phases, preventing phase wrapping after repetition. Energies are decoded without shifting them toward \(E_0\). Proposals with numerically unresolved evaluations are not accepted.

\newpage
\paragraph{Energy certification.}
\label{app:energy_witness}
Evaluating an existing PF over some dataset can happen on a classical or quantum computer. For this purpose, a common framework where operators can be defined and measured in either hardware is desired. Related methods estimate energies and certify their accuracy on quantum hardware~\cite{lin2022heisenberg,wang2025energycertification}. We can express the chemical-accuracy test as an expectation-value bound. This requires the reference energy $E_0$ which can be obtained with a high-fidelity QPE. Define \(H_{\mathcal P}\) to have the compiled circuit's eigenvectors and the energies decoded in Eq.~\eqref{eq:pf_energy_selection} as its eigenvalues.
For the selected eigenstate \(\rho_\star\),
\begin{equation}
W_{\mathcal P}=I-\frac{(H_{\mathcal P}-E_0I)^2}{\varepsilon_{\rm chem}^2},
\qquad
\operatorname{Tr}(W_{\mathcal P}\rho_\star)
=1-\frac{(E_{\mathcal P}-E_0)^2}{\varepsilon_{\rm chem}^2}.
\label{eq:energy_accuracy_witness}
\end{equation}
More generally, \(W_{\mathcal P}\le\Pi_{\rm chem}\), where
\(\Pi_{\rm chem}\) projects onto the chemical-accuracy window. This operator therefore bounds the probability of an outcome in that window from below. It also applies to signed reaction energies and tests energy accuracy rather than state fidelity. Using it on quantum hardware would require a measurement scheme with controlled cost and uncertainty.

Beyond exact classical calculations, verification needs other reference information. Fidelity witnesses, related to entanglement witnesses~\cite{horodecki1996separability,terhal2000witnesses}, bound overlap with a target state. For pure fermionic Gaussian targets, they require expectation values of quadratic Majorana operators~\cite{gluza2018fidelitywitnesses}. Pauli propagation~\cite{rall2019paulipropagation,rudolph2026paulipropagation} and Majorana propagation~\cite{miller2025majoranapropagation} track observables through a circuit and can reduce classical cost when the operator expansion remains manageable. State fidelity alone, however, does not control QPE phase error. Nested-commutator expectations estimate leading Trotter energy shifts~\cite{mehendale2025trottererrors,maxwell2026practicaltrotter}. Certifying accuracy from these estimates also requires control of reference-state, higher-order, and synthesis errors. Our experiments used the compiled-energy calculation above.

\paragraph{QPE-inclusive resource cost.}
\label{app:tgate_cost_derivations}
With per-evolution cost \(N_T\) from Eq.~\eqref{eq:main_pf_t_cost},
\(Q\) full-evolution calls per QPE attempt, and \(R_{95}\) independent
attempts sufficient for \(95\%\) success, our readout model gives
\begin{equation}
Q=\max\!\left\{1,\left\lceil\frac{8\pi}{t\epsilon_{\rm QPE}}\right\rceil\right\},
\quad
R_{95}=\left\lceil\frac{\log(0.05)}{\log(1-p_\star/2)}\right\rceil,
\quad
T_{95}=R_{95}Q N_T.
\label{eq:t95_objective}
\end{equation}
Here \(t=r\Delta t=1/\lambda\) is the total simulated evolution time, with \(\lambda=\sum_{j=1}^{M}|c_j|\) the sum of absolute nonidentity Pauli coefficients and \(\hbar=1\). We assigned the remaining budget \(\epsilon_{\rm QPE}=\varepsilon_{\rm chem}-|E_{\mathcal P}-E_0|\) to readout, requiring \(|\widehat E_{\mathcal P}-E_{\mathcal P}|\le\epsilon_{\rm QPE}\). The corresponding phase tolerance \(t\epsilon_{\rm QPE}\) determines \(Q\). Conditional on the selected energy outcome, QPE satisfies this tolerance with probability at least \(1/2\), giving the per-attempt success bound \(p_\star/2\) used to determine \(R_{95}\). We then compute \(E_{\mathcal P}\) classically and estimated the QPE \(\text{T}\)-gate cost from these precision and probability bounds.

We adopted the \(8\pi\) query factor and rotation-cost convention of Ref.~\cite{reiher2017reaction}. Its sign-controlled construction adds only controlled-NOT (CNOT) gates and preserves rotation counts when reversing all rotation signs implements the inverse evolution. We did not separately synthesize or verify controlled circuits for each proposal. Other decompositions can change absolute \(\text{T}\) counts. The estimate excludes state preparation, Clifford gates, error correction, and routing.

For \(m\) unique reaction species, write \(\Delta E=\sum_i\nu_iE_i\) and \(\Delta E_{\mathcal P}=\sum_i\nu_iE_{i,\mathcal P}\), with compiled error \(\epsilon_\Delta=|\Delta E_{\mathcal P}-\Delta E|\). When \(\epsilon_\Delta<\varepsilon_{\rm chem}\), species \(i\) receives QPE energy tolerance \(\epsilon_{{\rm QPE},i}=(\varepsilon_{\rm chem}-\epsilon_\Delta)/(m|\nu_i|)\) and failure probability at most \(0.05/m\). The finite-precision estimate is \(\widehat{\Delta E}_{\mathcal P}=\sum_i\nu_i\widehat E_{i,\mathcal P}\). Using each species' time and peak probability in Eq.~\eqref{eq:t95_objective}, with \(0.05/m\) in place of \(0.05\), gives \(T_{95}^{\rm rxn}=\sum_iR_iQ_iN_{T,i}\). The union bound then guarantees at least one successful readout for each species with probability at least \(95\%\). Stoichiometric weights enter the energy sum and precision requirements. Each distinct species contributes once to the circuit cost.

\subsection{Baseline construction}
\label{app:standard_pf_baseline}

\begin{figure}[!t]
\centering
\includegraphics[width=\linewidth]{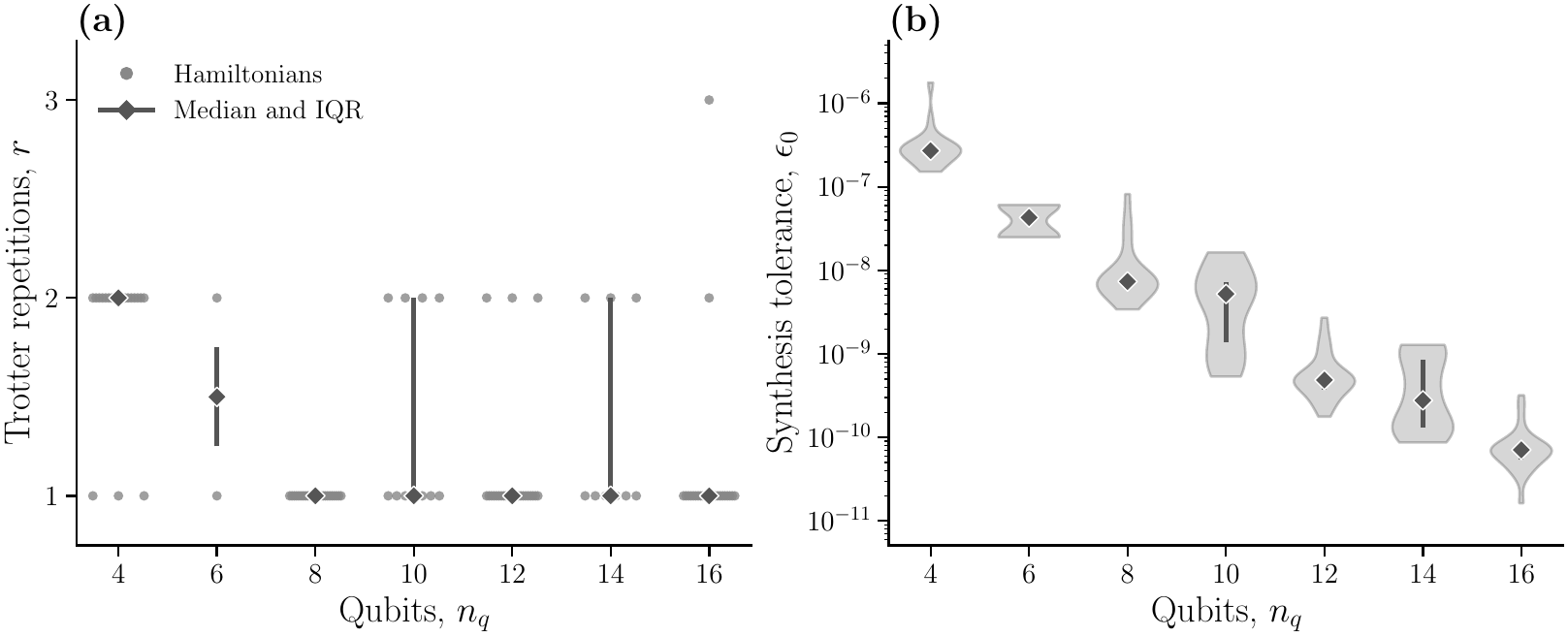}
\caption{Starting baseline parameters for the 115 training Hamiltonians spanning 4--16 qubits. The nested second-order tight baseline uses each Hamiltonian's evolution time \(t=1/\lambda\). (a) Repetition counts \(r\), with one point per Hamiltonian and horizontal offsets for visibility. (b) Distributions of uniform synthesis tolerance \(\epsilon_0\) at each qubit count, shown as violins on a logarithmic axis. Diamonds show medians and bars span the middle \(50\%\).}
\label{fig:baseline_initialization_vs_system_size}
\end{figure}

We chose the baseline parameters using the Trotter-error bound, error allocation, and cost model of Reiher \emph{et al.}~\cite{reiher2017reaction}, supplemental Eqs.~(19), (32), and (34). The loose baseline uses an analytical bound to estimate the required repetitions. The tight baseline uses a numerical energy calculation. Both retain all Hamiltonian contributions. Before starting the search, we synthesize the baseline rotations into Clifford+\(\text{T}\) gates and numerically verify that the resulting energy differs from the exact reference by less than \(\varepsilon_{\rm chem}\). Figure~\ref{fig:baseline_initialization_vs_system_size}
shows the starting parameters used in the 115-Hamiltonian search.

\newpage
\paragraph{PF and Trotter error.}
\label{app:baseline_hamiltonian_input}
The loose baseline orders Pauli strings lexicographically and bounds Trotter error using nonzero double commutators. Let
\(W_{\alpha\beta\alpha'}=\mathbf1([P_\alpha,[P_\beta,P_{\alpha'}]]\ne0)\).
The inner commutator is nonzero when \(P_\beta,P_{\alpha'}\) anticommute. The outer one requires \(P_\alpha\) to anticommute with exactly one of them. The bound coefficient is
\begin{equation}
B(H)=4\sum_{\alpha,\beta,\alpha'}
|c_\alpha c_\beta c_{\alpha'}|
\left(\mathbf1_{\alpha>\beta}\mathbf1_{\alpha'>\beta}
+\mathbf1_{\beta>\alpha}\mathbf1_{\alpha'=\alpha}\right)
W_{\alpha\beta\alpha'},
\label{eq:baseline_double_commutator_bound}
\end{equation}
We evaluated smaller sums directly and rounded upward. For larger sums, we sampled terms by coefficient magnitude, using \(5\times10^6\) samples for each of the two nonnegative sums and a one-sided \(1-10^{-9}\) confidence upper bound. This gives a Trotter scale without requiring an exact energy:
\begin{equation}
\epsilon_{\rm TS}(\Delta t)\le B(H)\Delta t^2,
\qquad
\beta(H)=\sqrt{B(H)/\epsilon_{\rm ref}},
\qquad
\epsilon_{\rm ref}=0.1~\mathrm{mHa}.
\label{eq:baseline_beta}
\end{equation}
For \(B(H)=0\), set \(\beta=1/t\). The flat second-order step is
\begin{equation}
S_2(\Delta t)=
\left(\prod_{j=1}^{M-1}e^{+ic_jP_j\Delta t/2}\right)
e^{+ic_MP_M\Delta t}
\left(\prod_{j=M-1}^{1}e^{+ic_jP_j\Delta t/2}\right).
\label{eq:baseline_strang_step}
\end{equation}
The full baseline evolution is \(U_0(t)=e^{+ic_0t}S_2(\Delta t)^r\).

\label{app:baseline_grouped_tight}
Our nested second-order baseline uses commuting layers of Hermitian one- and two-body fermionic interactions. We calibrated \(\beta(H)\) at a \(0.1\)~mHa Trotter-error target for each training Hamiltonian. For testing, we predicted it using $\log[\beta(H)t]\approx a_0+a_1\log n+a_2\log M+a_3\log\lambda$, fitted to the 115 training calibrations, where \(n\) counts qubits, \(M\) counts distinct nonidentity Pauli terms, and \(\lambda=\sum_j|c_j|\). The fit remained fixed during testing, and both baseline and ELOTE circuits were compiled and numerically checked.

\paragraph{Choosing repetitions and synthesis tolerance.}
\label{app:baseline_qpe_pf_scale}
We allocated the budget \(\epsilon_{\rm QPE}+\epsilon_{\rm TS}+\epsilon_{\rm Rot}\le\varepsilon_{\rm chem}\) to minimize the molecular cost model of Reiher \emph{et al.} Its cost constants are \(8\pi,4,11\), with synthesis bound \(C_{\rm synth}(\eta)\le4\log_2(1/\eta)+11\). Here, \(\mathcal M=M\) counts distinct Pauli strings, whereas \(\mathcal M=M_{\rm blk}\) for our nested baseline counts repeated contributions from different fermionic interactions separately.
Writing \(z=\epsilon_{\rm Rot}\), the minimum satisfies
\begin{equation}
3z\left[\ln\!\left(\frac{2\mathcal M\beta(H)}{z}\right)
+\frac{11\ln2}{4}\right]+2z-2\varepsilon_{\rm chem}=0,
\quad
\epsilon_{\rm TS}=\frac{\varepsilon_{\rm chem}-z}{3},
\quad
\epsilon_{\rm QPE}=\frac{2(\varepsilon_{\rm chem}-z)}{3}.
\label{eq:baseline_rotation_stationarity}
\end{equation}
We solve for \(z\) in \(0<z<\varepsilon_{\rm chem}\) by bisection. The allocation then sets the repetition count, step time, and common synthesis tolerance:
\begin{equation}
r=\max\!\left\{1,\left\lceil t\beta(H)
\sqrt{\frac{\epsilon_{\rm ref}}{\epsilon_{\rm TS}}}\right\rceil\right\},
\qquad \Delta t=t/r,
\qquad \epsilon_0\equiv\eta=\frac{\epsilon_{\rm Rot}\Delta t}{L_0}.
\label{eq:baseline_uniform_epsilon}
\end{equation}
\label{app:baseline_synthesis_precision}
Here, \(L_0\) is the number of rotations per second-order step. It equals \(2M-1\) when each distinct Pauli term is treated separately. For our nested baseline, we count rotations after combining consecutive rotations of the same Pauli string. The bound \(\epsilon_{\rm Rot}\le L_0\eta/\Delta t\) motivates the uniform baseline tolerance \(\eta=\epsilon_{\rm Rot}\Delta t/L_0\). Setting \(L=L_0\) and \(\epsilon_j=\eta\) in Eq.~\eqref{eq:main_pf_t_cost} gives the baseline \(\text{T}\)-gate count.

\FloatBarrier
\section{Information about the ELOTE}
\label{app:understanding_elote}

Here, we describe the ELOTE and \(\delta\)-ELOTE splitting and synthesis
rules, followed by search decisions and token use. We denote one ELOTE
step and full evolution by \(S_\star\) and \(U_\star\).

\subsection{Splitting, ordering, and coefficient aggregation}
\label{app:elote_splitting}

ELOTE adds contributions with matching Pauli strings and applies one rotation per string. We order the nonidentity strings as \(P_1<_{\rm lex}\cdots<_{\rm lex}P_M\), with \(I<X<Y<Z\), and retain the sign of each coefficient in its rotation angle. This gives the first-order Lie PF
\begin{equation}
S_\star(\Delta t)
=\prod_{j=1}^{M}e^{+ic_jP_j\Delta t},
\qquad
U_\star(t)=e^{+ic_0t}S_\star(t/r)^r.
\label{eq:elote_learned_step}
\end{equation}
ELOTE keeps the Jordan--Wigner representation and each Hamiltonian's baseline repetition count $r$ (one to three in training), replacing the nested second-order splitting in Appendix~\ref{app:baseline_grouped_tight}.

Our baseline implementation constructs rotations from the separate Jordan--Wigner expansions of Hermitian fermionic interaction terms. Different interactions can contribute to the same Pauli string. ELOTE sums these contributions before constructing its rotations. Writing each mapped interaction as \(H_g=\sum_P c_{gP}P\), the combined coefficient is \begin{equation} 
c_P=\sum_g c_{gP}. 
\label{eq:elote_coefficient_aggregation} 
\end{equation}
Contributions can cancel exactly, giving \(c_P=0\) and an identity rotation. Across one full evolution of each of the 115 training Hamiltonians, the baseline used 729,809 nonzero-angle rotations, compared with 106,854 for ELOTE. ELOTE also contained 59,924 zero-angle identities. All non-zero coefficients were kept.

\subsection{Mixed-precision synthesis rule}
\label{app:elote_precision}

Let \(c_{\rm ref}\) and \(W\) be the largest coefficient magnitude and Pauli weight among the separate Jordan--Wigner contributions from individual fermionic interactions, before contributions to identical Pauli strings are summed. Thus \(c_{\rm ref}\) can differ from the largest coefficient magnitude after summation. Define \(x_j=|c_j|/c_{\rm ref}\), and let \(w_j,z_j,q_j\) count the nonidentity, \(Z\), and \(X/Y\) letters in \(P_j\), respectively. With \(\epsilon_0\) the Hamiltonian's uniform baseline tolerance from Eq.~\eqref{eq:baseline_uniform_epsilon}, the rule is
\begin{equation}
\frac{\epsilon_j}{\epsilon_0}
=\left(1+1.2x_j^{0.1}\right)
 \left(1+0.12\frac{w_j-1}{W}\right)D_jG_H,
\label{eq:elote_tolerance_rule}
\end{equation}
where
\begin{equation}
D_j=
\begin{cases}
(1+1.2x_j)\left[1+0.18\dfrac{z_j}{W}(1+0.2x_j)\right],&q_j=0,\\[2pt]
1+0.3\dfrac{q_j}{W}(1+0.2x_j),&q_j>0,
\end{cases}
\qquad
G_H=1+0.12\left(\frac{c_{\rm ref}}{W\,(1~\mathrm{Ha})}\right)^{0.3}.
\label{eq:elote_character_factor}
\end{equation}
The Hartree scale in \(G_H\) fixes the empirical rule's units.
Tolerances \(\epsilon_j\) lie in \([10^{-15},1-10^{-15}]\), with no recorded value
reaching either bound. Holding other features fixed, larger coefficients or Pauli weights receive looser tolerances. Recorded ratios \(\epsilon_j/\eta\) ranged from 1.10 to 9.81. Rotation angle is not a separate input. The trend in Figure~\ref{fig:hamlib_pf_precision_reduction} follows from \(|\theta_j|=2|c_j|t/r\).

\newpage
\subsection{Reaction-specific \texorpdfstring{$\delta$}{delta}-ELOTE protocol}
\label{app:delta_elote_policy}

We initialized the reaction searches from an earlier molecular ELOTE protocol, before the final exponent changes in Eqs.~\eqref{eq:elote_tolerance_rule} and~\eqref{eq:elote_character_factor}. The selected \(\delta\)-ELOTE changes repetitions, ordering, and synthesis precision while keeping the signed Hamiltonian coefficients and Jordan--Wigner representation. Let $r_0(H)$ be the initial ELOTE repetition count and $m_H$ the number of nonidentity terms supplied to the protocol. Their mean absolute coefficient is $\bar c_H=\Lambda_{\ne I}(H)/m_H$, where $\Lambda_{\ne I}$ is the sum of their absolute coefficients. The new repetition count is
\begin{equation}
r_\delta(H)=\max\!\left\{1,
\left\lfloor r_0(H)\cdot\frac{\bar c_H}{\bar c_H+0.03~\mathrm{Ha}}
\right\rfloor\right\}.
\label{eq:delta_elote_repetitions}
\end{equation}
Terms are ordered by increasing Pauli weight, with diagonal ($I/Z$-only) terms first at each weight and Pauli labels breaking remaining ties. Matching strings are grouped and adjacent matching rotations can be combined. Distinct commuting strings remain separate rotations. Denoting this order by \(\sigma_\delta\), the first-order step is
\begin{equation}
S_\delta(\Delta t)=\prod_{j\in\sigma_\delta}
e^{+ic_jP_j\Delta t},\qquad
U_\delta(t)=e^{+ic_0t}S_\delta(t/r_\delta)^{r_\delta}.
\label{eq:delta_elote_step}
\end{equation}
The synthesis rule uses the same $x_j$, $w_j$, $q_j$, $z_j$, and $W$ as above, with $c_{\rm ref},W>0$ for the recorded Hamiltonians:
\begin{equation}
\frac{\epsilon^{\delta}_j}{\epsilon_0}
=\left(1+1.2x_j^{0.3}\right)
 \left(1+0.13\frac{w_j-1}{W}\right)D^{\delta}_jG^{\delta}_H,
\label{eq:delta_elote_tolerance}
\end{equation}
where
\begin{equation}
D^{\delta}_j=
\begin{cases}
(1+1.2x_j)\left[1+0.30\dfrac{z_j}{W}(1+0.2x_j)\right],&q_j=0,\\[2pt]
1+0.3\dfrac{q_j}{W}(1+0.2x_j),&q_j>0,
\end{cases}
\qquad
G^{\delta}_H=1+0.250
\left(\frac{c_{\rm ref}}{W\,(1~\mathrm{Ha})}\right)^{0.4}.
\label{eq:delta_elote_factors}
\end{equation}
The tolerance is restricted to $[10^{-15},1-10^{-15}]$. The selected protocol met chemical accuracy for all ten reactions, with a maximum error of $0.309$~mHa. Repetitions, ordering, and precision changed together, so the measured savings do not isolate their individual contributions or provide an analytical error guarantee.

\FloatBarrier
\subsection{Search decisions and inference effort}
\label{app:elote_seed_effort}

Across the five searches, 13\%--27\% of proposals lowered the cost over the training ensemble while satisfying all 115 energy constraints (Table~\ref{tab:elote_seed_effort}). Iteration 0 is excluded.

\begin{table}[!ht]
\centering
\small
\setlength{\tabcolsep}{5pt}
\begin{tabular}{cccccc}
\toprule
Search & Rounds & Accepted (\%) & Rejected (\%) &
\begin{tabular}[c]{@{}c@{}}Input tokens/round\\(mean $\pm$ SD)\end{tabular} &
\begin{tabular}[c]{@{}c@{}}Output tokens/round\\(mean $\pm$ SD)\end{tabular} \\
\midrule
1 & 100 & 13.0 & 87.0 & $(0.2880 \pm 0.2423)\,\mathrm{M}$ & $(0.0210 \pm 0.0166)\,\mathrm{M}$ \\
2 & 100 & 24.0 & 76.0 & $(0.2658 \pm 0.2184)\,\mathrm{M}$ & $(0.0191 \pm 0.0127)\,\mathrm{M}$ \\
3 & 100 & 26.0 & 74.0 & $(0.2929 \pm 0.2222)\,\mathrm{M}$ & $(0.0193 \pm 0.0115)\,\mathrm{M}$ \\
4 & 100 & 24.0 & 76.0 & $(0.2471 \pm 0.1877)\,\mathrm{M}$ & $(0.0168 \pm 0.0105)\,\mathrm{M}$ \\
5 & 100 & 27.0 & 73.0 & $(0.3037 \pm 0.2283)\,\mathrm{M}$ & $(0.0175 \pm 0.0123)\,\mathrm{M}$ \\
\bottomrule
\end{tabular}
\caption{Search decisions and token use per mutation. Acceptance and rejection percentages use all 100 rounds per search. Token means and standard deviations use only rounds with complete token logs. Counts include preparation, editing, retries, and repeated input context. M denotes million tokens. Rejected proposals include those without a conclusive evaluation and do not necessarily violate chemical accuracy.}
\label{tab:elote_seed_effort}
\end{table}

Of 500 proposals, 114 improved cost, 181 did not, and 70 violated the QPE energy constraint. The other 135 had inconclusive numerical checks (107), execution errors (26), or both (2). Figure~\ref{fig:hamlib_pf_search_trajectories}a omits these 135 proposals and another 15 whose QPE energy-constraint violations were confirmed before complete cost scoring. For those 15, a probability bound ruled out a chemically accurate QPE peak with probability above $1/2$ for at least one Hamiltonian. Numerical uncertainty does not establish physical inaccuracy. The token statistics cover 644 agent attempts, including 144 retries, across the same searches and steps.
\FloatBarrier

\end{document}